\documentclass{ws-rv9x6}
\usepackage{subfigure}   
\usepackage{ws-rv-thm}   
\usepackage{hyperref}
\usepackage{xcolor}
\hypersetup{
    colorlinks,
    linkcolor={red!50!black},
    citecolor={blue!50!black},
    urlcolor={blue!80!black}
}
\usepackage{mdframed}
\usepackage{mathtools}
\usepackage{dsfont}
\usepackage{booktabs}
\usepackage{algorithmic}
\usepackage{algorithm}

\usepackage{tikz}
\usetikzlibrary{arrows.meta, calc, chains, positioning}
\usepackage{pgfplots}
\usepackage{color,soul}

\tikzstyle{neuron}=[draw,circle,minimum size=20pt,inner sep=0pt, fill=white]
\tikzstyle{stateTransition}=[thick]
\tikzstyle{learned}=[text=black]

\usepackage{ws-rv-van}   
\makeindex

\usepackage[strict]{changepage}

\usepackage{framed}

\definecolor{formalshade}{rgb}{0.95,0.95,1}
\definecolor{darkblue}{rgb}{0.0, 0.0, 0.55}
\definecolor{tabblue}{RGB}{31, 119, 180}
\definecolor{tabred}{RGB}{214, 39, 40}

\def\eg{\textit{e.g.}}
\def\ie{\textit{i.e.}}
\newcommand{\etal}{\textit{et al}.}

\newenvironment{formal}{%
  \MakeFramed{\advance\hsize-\width\FrameRestore}%
  \noindent\hspace{-4.55pt}
  \begin{adjustwidth}{}{7pt}%
  \vspace{2pt}\vspace{2pt}%
}
{%
  \vspace{2pt}\end{adjustwidth}\endMakeFramed%
}

\pgfkeys{/pgfplots/tuftelike/.style={
  semithick,
  tick style={major tick length=4pt,semithick,black},
  separate axis lines,
  axis x line*=bottom,
  axis x line shift=10pt,
  xlabel shift=10pt,
  axis y line*=left,
  axis y line shift=10pt,
  ylabel shift=10pt}}

\begin{document}
\newcommand\n{5}
\newcommand\nbm{8}

\chapter[Statistical  Mechanics  and  Artificial Neural Networks]{Statistical Mechanics and  Artificial Neural Networks: Principles, Models, and Applications\label{ra_ch4}}

\author[L. B\"ottcher and G. Wheeler]{Lucas B\"ottcher\footnote{Lucas B\"ottcher is also with the Dept.\ of Medicine, University of Florida, Gainesville, FL,
32610, United States of America.} and Gregory Wheeler}

\address{Dept.\ of Computational Science and Philosophy, Frankfurt School of Finance and Management, Adickesallee 32--34, 60322 Frankfurt am Main, Germany \\
\href{mailto:l.boettcher@fs.de}{l.boettcher@fs.de} and \href{mailto:g.wheeler@fs.de}{g.wheeler@fs.de}}

\begin{abstract}
The field of neuroscience and the development of artificial neural networks (ANNs) have mutually influenced each other, drawing from and contributing to many concepts initially developed in statistical mechanics. Notably, Hopfield networks and Boltzmann machines are versions of the Ising model, a model extensively studied in statistical mechanics for over a century. In the first part of this chapter, we provide an overview of the principles, models, and applications of ANNs, highlighting their connections to statistical mechanics and statistical learning theory. 

Artificial neural networks can be seen as high-dimensional mathematical functions, and understanding the geometric properties of their loss landscapes (\ie, the high-dimensional space on which one wishes to find extrema or saddles) can provide valuable insights into their optimization behavior, generalization abilities, and overall performance. Visualizing these functions can help us design better optimization methods and improve their generalization abilities. Thus, the second part of this chapter focuses on quantifying geometric properties and visualizing loss functions associated with deep ANNs.
\end{abstract}


\body

\tableofcontents
\section{Introduction}
\label{sec:intro}
There has been a long tradition in studying learning systems within statistical mechanics.\cite{zdeborova2016statistical,carleo2019machine,mehta2019high} Some versions of artificial neural networks (ANNs) were, in fact, inspired by the Ising model that has been proposed in 1920 by Wilhelm Lenz as a model for ferromagnetism and solved in one dimension by his doctoral student, Ernst Ising, in 1924.\cite{amit1985spin,ising2017fate,folk2023survival} In the language of machine learning, the Ising model can be considered as a non-learning recurrent neural network (RNN).

Contributions to ANN research extend beyond statistical mechanics, however, requiring a broad range of scientific disciplines.  Historically, these efforts have been marked by eras of interdisciplinary collaboration referred to as cybernetics\cite{rosenblueth1943behavior,wiener2019cybernetics}, connectionism\cite{thorndike1932fundamentals,berkeley2019curious}, artificial intelligence\cite{minsky1972perceptrons,RussellNorvig:2009}, and machine learning\cite{VC:1974,Mitchell:1997}.  To emphasize the importance of multidisciplinary contributions in ANN research, we will often highlight the research disciplines of individuals who have played pivotal roles in the advancement of ANNs.

\subsection{The McCulloch–Pitts calculus and Hebbian learning}

An influential paper by the neuroscientist Warren S.~McCulloch and logician Walter Pitts published in 1943 is often regarded as inaugurating neural network research,\cite{mcculloch1943logical} but in fact there was an active research community at that time working on the mathematics of neural activity.\cite{Abraham:2004}  McCulloch and Pitts's pivotal contribution was to join propositional logic and Alan Turing's mathematical notion of computation\cite{Turing:1937} to propose a calculus for creating compositional representations of neural behavior. 
Their work has later on been picked up by the mathematician Stephen C.\ Kleene\cite{kleene1956representation}, who made some of their work more accessible.\footnote{Kleene comments on the work by McCulloch and Pitts in his 1956 article\cite{kleene1956representation}: ``The present memorandum is partly an exposition of the McCulloch–Pitts results; but we found the part of their
paper which treats of arbitrary nets obscure; so we have proceeded independently here.''}

Contemporaneously, the physchologist Donald O.\ Hebb formulated in his 1949 book ``The Organization of Behavior''\cite{hebb1949organization} a neurophysiological postulate, which is known today as ``Hebb's rule'' or ``Hebbian learning'':
\begin{formal}
Let us assume then that the persistence or repetition of a reverberatory activity (or ``trace'') tends to induce lasting cellular changes that add to its stability. ... When an axon of cell $A$ is near enough to excite a cell $B$ and repeatedly or persistently takes part in firing it, some growth process or metabolic change takes place in one or both cells such that $A$'s efficiency, as one of the cells firing $B$, is increased.
\vspace{1em}\\
\footnotesize{Donald O.\ Hebb in ``The Organization of Behavior'' (1949)}
\end{formal}
\noindent In the modern literature, Hebbian learning is often summarized by the slogan ``neurons wire together if they fire together''.\cite{lowel1992selection}
\subsection{The Perceptron}
In 1958, the psychologist Frank Rosenblatt developed the perceptron\cite{rosenblatt1958perceptron,rosenblatt1962principles}, which is essentially a binary classifier that is based on a single artificial neuron and a Hebbian-type learning rule.\footnote{The artificial neuron of this type is sometimes referred to as ``McCulloch–Pitts neuron'', owing to its connection to the foundational work of McCulloch and Pitts \cite{mcculloch1943logical}. Some authors distinguish McCulloch–Pitts neurons from perceptrons by noting that the former handle binary signals, while the latter can process real-valued inputs.} In parallel to Rosenblatt's work, the electrical engineers Bernard Widrow and Ted Hoff developed a similar ANN named ADALINE (Adaptive Linear Neuron).\cite{widrow1960adaptive}

\begin{figure}[t]
\centering
\begin{tikzpicture}[
    node distance = 10mm and 25mm,
    start chain = going below,
    arro/.style = {-Latex},
    bloque/.style = {text width=4ex, inner sep=1pt, align=right, on chain},
]
\node[bloque] (in-1) {$x_{1}$};
\node[bloque] (in-2) {$x_{2}$};
\node[bloque] (in-3) {$\dots$};
\node[bloque] (in-4) {$x_{n}$};

\node (out) [circle, draw=black, fill=blue!20, minimum size=14mm,
right=of $(in-2)!0.5!(in-3)$]  {$\Sigma$};

\foreach \i in {1,2,4}
    \draw[arro] (in-\i) -- (out);

\draw (in-1) -- (out) node[midway, sloped, above] {$w_{1}$};
\draw (in-2) -- (out) node[midway, sloped, above] {$w_{2}$};
\draw (in-3) -- (out) node[midway, sloped, above] {$\dots$};
\draw (in-4) -- (out) node[midway, sloped, above] {$w_{n}$};

\draw[arro] ($(out.north)+(0,1.5)$) -- ($(out.south)+(0,1.4)$) node[midway, left] {$b$};

\node (threshold) [circle, draw=black, fill=blue!20, minimum size=14mm, right=of out.west, xshift=-1mm]  {};
\begin{axis}[
    at=(threshold),
    anchor=center,
    hide axis,
    ymin=0, ymax=1,
    xmin=-3, xmax=3,
    width=0.8cm,
    height=0.8cm,
    scale only axis,
    samples=500,
    domain=-3:3,
    line width=0.5pt
    ]
    \addplot[black] {ifthenelse(x > 0, 1, 0)};
\end{axis}

\draw[arro] (out) -- (threshold) node[midway, above] {};

\coordinate[right=of threshold, xshift=-8mm] (output);
\draw[arro] (threshold) -- (output) node[right] {$y$};

\end{tikzpicture}
\caption{A schematic of a perceptron with output $y$, inputs $x_1, x_2, \dots, x_n$, corresponding weights $w_1, w_2, \dots, w_n$, and a bias term $b$. Traditionally, the inputs and outputs of perceptrons were considered to be binary (\eg, $\{-1, 1\}$ or $\{0, 1\}$), while weights and biases are generally represented as real numbers. The symbol $\Sigma$ signifies the summation process wherein the artificial neuron computes $\sum_{i}w_i x_i + b$. This computed value then becomes the input for a step activation function, denoted by a step symbol. We denote the final output as $y$. While the original perceptron employed a step activation function, contemporary applications frequently utilize a variety of functions including sigmoid, $\tanh$, and rectified linear unit (ReLU).}
\label{fig:neuron}
\end{figure}
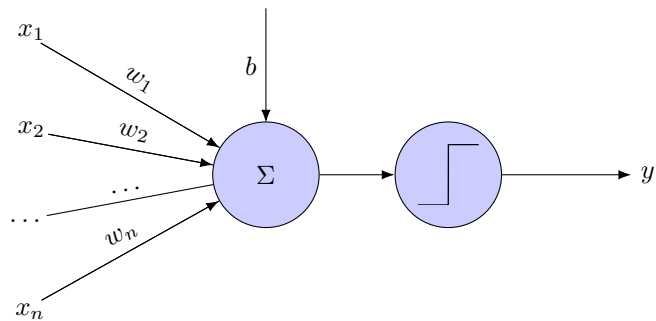

We show a schematic of a perceptron in Fig.~\ref{fig:neuron}. It receives inputs $x_1,x_2,\dots,x_n$ that are weighted with $w_1,w_2,\dots,w_n$. Within the summer (indicated by a $\Sigma$ symbol), the artificial neuron computes $\sum_i w_i x_i +b$, where $b$ is a bias term. This quantity is then used as input of a step activation function (indicated by a step symbol). The output of this function is $y$. Such simple artificial neurons can be used to implement logical functions including \texttt{AND}, \texttt{OR}, and \texttt{NOT}. For example, consider a perceptron with Heaviside activation, two binary inputs $x_1,x_2\in\{0,1\}$, weights $w_1,w_1=1$ and bias $b=-1.5$. We can readily verify that $y=x_1\wedge x_2$. An example of a logical function that cannot be represented by a perceptron is \texttt{XOR}. This was shown by the computer scientists Marvin L.\ Minsky and Seymour A.\ Papert in their 1969 book ``Perceptrons: An Introduction to Computational Geometry''\cite{minsky1972perceptrons}. Based on their analysis of perceptrons they considered the extension of research on this topic to be ``sterile''. 

Problems like the one associated with representing \texttt{XOR} using a single perceptron can be overcome by employing multilayer perceptrons (MLPs) or, in a broader sense, multilayer feedforward networks equipped with various types of activation functions.\cite{amari1967theory,ivakhnenko1967cybernetics,ivakhnenko1971polynomial} One effective way to train such multilayered structures is reverse mode automatic differentiation (\ie, backpropagation).\cite{kelley1960gradient,linnainmaa1970representation,linnainmaa1976taylor,werbos1982applications,rumelhart1986learning}

Notice that perceptrons and other ANNs predominantly operate on real-valued data. Nevertheless, there is an expanding body of research focused on ANNs that are designed to handle complex-valued and, in some cases, quaternion-based information.\cite{noest1987,noest1988discrete,noest1988associative,kobayashi2010exceptional,minemoto2016quaternionic,zhang2021optical,bottcher2022complex}

\subsection{Objective Bayes, entropy, and information theory}
Over the same period of time, seminal work in probability and statistics appeared that would later prove crucial to the development of ANNs.  In the early 20th century, a renewed interest in inverse ``Bayesian`` inference \cite{Bayes:1763,Laplace:1774}  met strong resistance from the emerging frequentist school of statistics led by Jerzy Neyman, Karl Pearson, and especially  Ronald Fisher, who advocated his own {\em fiducial inference}  as a better alternative to Bayesian inference \cite{Seidenfeld:1992,Seidenfeld:2007}. In response to criticism that Bayesian methods were irredeemably subjective, Harold Jeffreys presented a methodology for ``objectively'' assigning prior probabilities based on symmetry principles and invariance properties.\cite{Jeffreys:1939} 

Edwin Jaynes, building on Jeffreys's ideas for objective Bayesian priors and Shannon's quantification of uncertainty\cite{Shannon:1948}, proposed the Maximum Entropy (MaxEnt) principle as a rationality criteria for assigning prior probabilities that are at once consistent with known information while remaining maximally uncommitted about what remains uncertain\cite{jaynes1957information,jaynes1957informationii,Jaynes:1973}.  
Jaynes envisioned the MaxEnt principle as the basis for a ``logic of science''\cite{Jaynes:2003}. For instance, he proposed a reinterpretation of the principles of thermodynamics in information-theoretic terms where macro-states of a physical system are understood through the partial information they provide about the possible underlying microstates, showing how the original Boltzmann–Gibbs interpretation of entropy could be replaced by a more general single framework based on Shannon entropy.  
Objective Bayesianism does not remove subjective judgment \cite{Seidenfeld_ObjectiveBayes:1979}, MaxEnt is far from a universal principle for encoding prior information\cite{Wheeler.calibration}, and the terms `objective' and `subjective' are outdated.\cite{Hacking:2015,Gelman:2017} 

For example, the MaxEnt principle  can be used to select the least biased distribution from  many common exponential family distributions by imposing constraints that match the expectation values of their sufficient statistics.  That is, from constraints on the expected values of some function $h(x)$ on data parameterized by $\theta$, applying the MaxEnt principle to find the probability distribution that maximizes entropy under those constraints often takes the form of an exponential family distribution  
\begin{equation}
    p(x \mid \theta) = h(x) \, \exp\!\left( \theta^{\top} T(x) - A(\theta) \right)\,,
\end{equation}
\noindent where $h(x)$ is the underlying base measure, $\theta$ is the natural parameter of the distribution, $A(\theta)$ is the log normalizer, and $T(x)$ is the sufficient statistic determined by MaxEnt under the given constraints.   MaxEnt derived distributions depend on contingencies of the data and the availability of the right constraints to ensure the derivation goes through, however--contingencies that undermine viewing MaxEnt as a fundamental principle of rational inference.

Nevertheless, Jaynes's  notion of treating probability as an extension of logic\cite{Jaynes:2003,Jaynes:2003,progicnet.fotfs} and probabilistic inference  as governed by coherence conditions on information states\cite{deCooman:2005,progicnet} has proved useful in optimization (\eg, cross-entropy loss \cite{Cox:1958}), entropy-based regularization of ANNs\cite{Hinton_vanCamp:1993}, and Bayesian neural networks.\cite{Neal:1996} 
\subsection{Connections to the Ising model}
Returning to the Ising model, a modified version of it has been endowed with a learning algorithm by the mathematical neuroscientist Shun'ichi Amari in 1972 and has been employed for learning patterns and sequences of patterns.\cite{amari1972learning} Amari's self-organizing ANN is an early model of associative memory that has later been popularized by the physicist John Hopfield.\cite{hopfield1982neural} Despite its limited representational capacity and use in practical applications, the study of Hopfield-type learning systems is still an active research area.\cite{isokawa2008associative,minemoto2016quaternionic,ramsauer2020hopfield,benedetti2022supervised}

The Hopfield network shares the Hamiltonian with the Ising model. Its learning algorithm is deterministic and aims at minimizing the ``energy'' of the system. One problem with the deterministic learning algorithm of Hopfield networks is that it cannot escape local minima. A solution to this problem is to employ a stochastic learning rule akin to Monte-Carlo methods that are used to generate Ising configurations.\cite{bottcher2021computational} In 1985, David H.\ Ackley, Geoffrey E.\ Hinton, and Terrence J.\ Sejnowski proposed such a learning algorithm for an Ising-type ANN which they called ``Boltzmann machine'' (BM).\cite{DBLP:conf/aaai/FahlmanHS83,hinton1983analyzing,hinton1983optimal,ackley1985learning} In 1986, Smolensky introduced the concept of ``Harmonium''\cite{smolensky1986}, which later evolved into restricted Boltzmann machines (RBMs). In contrast to Boltzmann machines, RBMs benefit from more efficient training algorithms that were developed in the early 2000s.\cite{hinton2002training,hinton2006fast} Since then, RBMs have been applied in various context such as to reduce dimensionality of datasets\cite{hinton2006reducing}, study phase transitions\cite{kim2018smallest,efthymiou2019super,mehta2019high,d2020learning},
represent wave functions\cite{carleo2017solving,torlai2018neural}.

Our historical overview of ANNs aims at emphasizing the strong interplay between statistical mechanics and related fields in the investigation of learning systems. Given the extensive span of over eight decades since the pioneering work of McCulloch and Pitts, summarizing the history of ANNs within a few pages cannot do justice to its depth and significance. We therefore refer the reader to the excellent review by J\"urgen Schmidhuber (see Ref.~\refcite{schmidhuber2015deep}) for a more detailed overview of the history of deep learning in ANNs.

As computing power continues to increase, deep (\ie, multilayer) ANNs have, over the past decade, become pervasive across various scientific domains.\cite{dean2022golden} Different established and newly developed ANN architectures have not only been employed to tackle scientific problems but have also found extensive applications in tasks such as image classification and natural language processing.\cite{hochreiter1997long,DBLP:journals/pieee/LeCunBBH98,DBLP:conf/nips/GoodfellowPMXWOCB14,DBLP:conf/nips/VaswaniSPUJGKP17} 

Finally, we would also like to emphasize an area of research that has significantly advanced due to the availability of automatic differentiation\cite{paszke2017automatic} and the utilization of ANNs as universal function approximators.\cite{cybenko1989approximation,hornik1989multilayer,hornik1991approximation} This area resides at the intersection of non-linear dynamics\cite{wang1998runge,DBLP:conf/nips/ChenRBD18}, control theory\cite{kaiser2021data,schafer2021control,bottcher2022ai,asikis2022neural,bottcher2022near,bottcher2023control,bottcher2023gradient,ruiz2023neural,asikis2021multi}, and dynamics-informed learning, where researchers often aim to integrate ANNs and their associated learning algorithms with domain-specific knowledge.\cite{raissi2019physics,brunton2022data,mowlavi2023optimal,fronk2023interpretable,xia2023spectrally,bottcher2024control} This integration introduces an inductive bias that facilitates effective learning.

The outline of this chapter is as follows. To better illustrate the connections between the Ising model and ANNs, we devote Secs.~\ref{sec:hopfield} and \ref{sec:bm} to the Hopfield model and Boltzmann machines, respectively. In Sec.~\ref{sec:loss_landscapes}, we discuss how mathematical tools from high-dimensional probability and differential geometry can help us study the loss landscape of deep ANNs.\cite{bottcher2024visualizing} We conclude this chapter in Sec.~\ref{sec:conclusions}.
\section{Hopfield network}
\label{sec:hopfield}
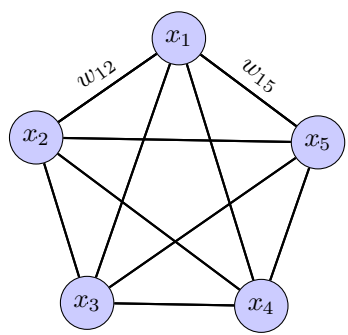
\begin{figure}
\centering
\begin{tikzpicture}[scale=1.3]
    \begin{scope}[rotate=17]
        \foreach \number in {1,...,\n}{
            \node[neuron, draw=black, fill=blue!20] (N-\number) at ({\number*(360/\n)}:1.5cm) {$x_\number$};
        }

        \foreach \number in {1,...,\n}{
            \foreach \y in {1,...,\n}{
                \draw[stateTransition] (N-\number) -- (N-\y);
            }
        }
        \end{scope}
        \begin{scope}[rotate=-1]
        \draw[learned,stateTransition] (N-1) -- (N-2) node [midway,above=0cm,sloped] {$w_{12}$};
        \draw[learned,stateTransition] (N-1) -- (N-5) node [midway,above=0cm,sloped] {$w_{15}$};
    \end{scope}
\end{tikzpicture}
\caption{An example of a Hopfield network with $n=5$ neurons. Each neuron is connected to all other neurons through black edges, representing both inputs and outputs. (Adapted from \citen{martin_thoma_github}.)}
\label{fig:hopfield_net}
\end{figure}

A Hopfield network is a complete (\ie, fully connected) undirected graph in which each node is an artificial neuron of McCulloch--Pitts type (see Fig.~\ref{fig:neuron}). We show an example of a Hopfield network with $n=5$ neurons in Fig.~\ref{fig:hopfield_net}. We use $x_i\in\{-1,1\}$ to denote the state of neuron $i$ ($i\in\{1,\dots,n\}$). Because the underlying graph is fully connected, neuron $i$ receives $n-1$ inputs $x_j$ ($j\neq i$). The inputs $x_j$ associated with neuron $i$ are assigned weights $w_{ij}\in\mathbb{R}$. In a Hopfield network, weights are assumed to be symmetric (\ie, $w_{ij}=w_{ji}$), and self-weights are considered to be absent (\ie, $w_{ii}=0$). 

With these definitions in place, the activation of neuron $i$ is given by
\begin{equation}
a_i=\sum_j w_{i j} x_j+b_i\,,
\label{eq:activation}
\end{equation}
where $b_i$ is the bias term of neuron $i$.

Regarding connections between the Ising model and Hopfield networks, the states of artificial neurons align with the binary values employed in the Ising model to model the orientations of elementary magnets, often referred to as classical ``spins''. The weights in a Hopfield network represent the potentially diverse couplings between different elementary magnets. Lastly, the bias terms assume the function of external magnetic fields that act on these elementary magnets. Given these structural similarities between the Ising model and  Hopfield networks, it is natural to assign them the Ising-type energy function
\begin{equation}
E=-\frac{1}{2}\sum_{i,j}w_{i j} x_i x_j - \sum_i b_i x_i\,.
\label{eq:energy_hopfield}
\end{equation}

Hopfield networks are not just a collection of interconnected artificial neurons, but they are dynamical systems whose states $x_i$ evolve in discrete time according to
\begin{equation}
x_i \leftarrow \begin{cases}
  1\,, & \text{if }a_i\geq 0\\
  -1\,, & \text{otherwise}\,,
\end{cases}
\label{eq:hopfield_update}
\end{equation}
where $a_i$ is the activation of neuron $i$ [see Eq.~\eqref{eq:activation}].

For asynchronous updates in which the state of one neuron is updated at a time, the update rule \eqref{eq:hopfield_update} has an interesting property: Under this update rule, the energy $E$ [see Eq.~\eqref{eq:energy_hopfield}] of a Hopfield network never increases.

The proof of this statement is straightforward. We are interested in the cases where an update of $x_i$ causes this quantity to change its sign. Otherwise the energy will stay constant. Consider the two cases (i) $a_i=\sum_j w_{i j} x_j +b_i<0$ with $x_i=1$, and (ii) $a_i\geq 0$ with $x_i=-1$. In the first case, the energy difference is
\begin{equation}
\Delta E_i = E(x_i=-1)-E(x_i=1)= 2  \Bigl(b_i + \sum_{j} w_{i j} x_j\Bigr)=2 a_i\,.
\label{eq:energy_difference_p1_m1}
\end{equation}
Notice that $\Delta E_i < 0$ because $a_i=\sum_j w_{i j} x_j +b_i< 0$.

In the second case, we have
\begin{equation}
\Delta E_i = E(x_i=1)-E(x_i=-1)= -2 \Bigl( b_i + \sum_{j} w_{i j} x_j\Bigr) =-2 a_i\,.
\label{eq:energy_difference_m1_p1}
\end{equation}
The energy difference satisfies $\Delta E_i\leq 0$ because $a_i\geq 0$. In summary, we have shown that $\Delta E_i\leq 0$ for all changes of the state variable $x_i$ according to update rule \eqref{eq:hopfield_update}.

Equations \eqref{eq:energy_difference_p1_m1} and \eqref{eq:energy_difference_m1_p1} also show that the energy difference associated with a change of sign in $x_i$ is $2 a_i$ and $-2 a_i$ for $a_i<0$ and $a_i\geq 0$, respectively. Absorbing the bias term $b_i$ in the weights $w_{ij}$ by associating an extra active unit to every node in the network yields for the corresponding energy differences
\begin{equation}
\Delta E_i = \pm 2 \sum_{j} w_{i j} x_j\,.
\label{eq:energy_difference2}
\end{equation}

One potential application of Hopfield networks is to store and retrieve information in local minima by adjusting their weights. For instance, consider the task of storing $N$ binary (black and white) patterns in a Hopfield network. Let these $N$ patterns be represented by the set $\{p_i^{\nu}=\pm 1 | 1 \leq i \leq n \}$ with $\nu \in \{1,\dots,N\}$.

To learn the weights that represent our binary patterns, we can apply the Hebbian-type rule

\begin{equation}
w_{i j} = w_{j i} = \frac{1}{N}\sum_{\nu=1}^N p_{i}^\nu p_{j}^\nu\,.
\label{eq:hebb_hopfield}
\end{equation}

In Hopfield networks with a large number of neurons, the capacity for storing and retrieving patterns is limited to approximately 14\% of the total number of neurons.\cite{hertz1991introduction} However, it is possible to enhance this capacity through the use of modified versions of the learning rule \eqref{eq:hebb_hopfield}.\cite{DBLP:conf/icann/Storkey97}

After applying the Hebbian learning rule to adjust the weights of a given Hopfield network, we may be interested in studying the evolution of different initial configurations $\{x_i\}$ according to update rule \eqref{eq:hopfield_update}. A Hopfield network accurately represents a pattern $\{p_i^{\nu}=\pm 1| 1 \leq i \leq n \}$ if $x_i$ remains equal to $p_i^\nu$ both before and after an update for all $i$ (meaning $\{p_i^\nu\}$ is a fixed point of the system). Depending on the number of stored patterns and the chosen initial configuration, the Hopfield network may converge to a local minimum that does not align with the desired pattern. To mitigate this behavior, one can employ a stochastic update rule instead of the deterministic rule \eqref{eq:hopfield_update}. This will be the topic that we are going to discuss in the following section.
\section{Boltzmann machine learning}
\label{sec:bm}
In Hopfield networks, if we start with an initial configuration that is close enough to a desired local energy minimum, we can recover the corresponding pattern $\{p_i^\nu\}$. However, for certain applications, relying solely on the deterministic update rule \eqref{eq:hopfield_update} may not be enough as it never accepts transitions that are associated with an increase in energy. In constraint satisfaction tasks, we often require learning algorithms to have the capability to occasionally accept transitions to configurations of higher energy and move the system under consideration away from local minima towards a global minimum. A method that is commonly used in this context is the M(RT)$^2$ algorithm introduced by Nicholas Metropolis, Arianna W.\ Rosenbluth, Marshall N. Rosenbluth, Augusta H. Teller, and Edward Teller in their seminal 1953 paper ``Equation of State Calculations by Fast Computing Machines''.\cite{metropolis1953equation,gubernatis2005marshall} This algorithm became the basis of many optimization methods, such as simulated annealing.\cite{kirkpatrick1983optimization}

In the 1980s, the M(RT)$^2$ algorithm has been adopted to equip Hopfield-type systems with a stochastic update rule, in which neuron $i$ is activated (set to $1$) regardless of its current state, with probability
\begin{equation}
\sigma_i \equiv \sigma (\Delta E_i/T) = \sigma(2 a_i/T)=\frac{1}{1+\exp(-\Delta E_i/T)}\,.
\label{eq:boltzmann_machine}
\end{equation}
Otherwise, it is set to $-1$. The corresponding ANNs were dubbed ``'Boltzmann machines''.\cite{DBLP:conf/aaai/FahlmanHS83,hinton1983analyzing,hinton1983optimal,ackley1985learning} 
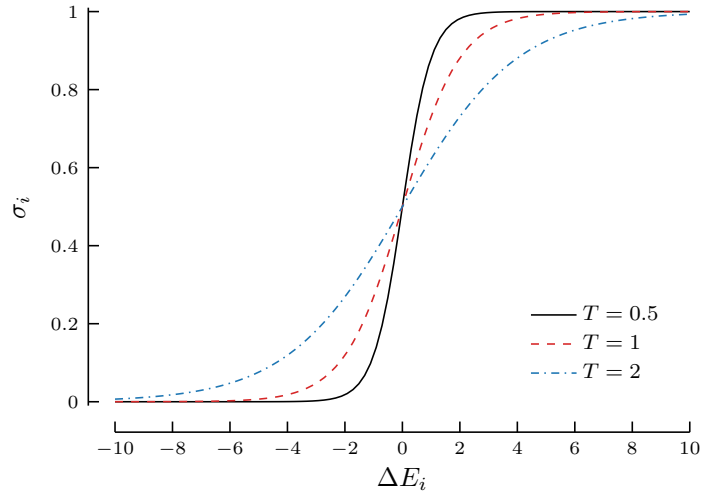
\begin{figure}[t]
\centering
\begin{tikzpicture}[every pin/.style={red!50!black,font=\small\sffamily}]
\begin{axis}[
    tuftelike,
    width=0.8\textwidth, 
    height=0.6\textwidth, 
    xlabel={$\Delta E_i$},
    ylabel={$\sigma_i$},
    xlabel style={yshift=-0.5em}, 
    legend pos=south east,
    legend cell align=left, 
    legend style={draw=none},
    ymin=-0.01,
    ymax=1.01,
    xmin=-10,
    xmax=10,
    enlarge x limits=false,
    axis line style={-}, 
    ticklabel style={font=\scriptsize}, 
    legend style={font=\footnotesize}, 
]

\addplot[black, domain=-10:10, samples=100] {1/(1 + exp(-x/0.5))};
\addlegendentry{$T = 0.5$};

\addplot[tabred, domain=-10:10, samples=100, dashed] {1/(1 + exp(-x/1))};
\addlegendentry{$T = 1$};

\addplot[tabblue, domain=-10:10, samples=100, dashdotted] {1/(1 + exp(-x/2))};
\addlegendentry{$T = 2$};

\end{axis}
\end{tikzpicture}
\caption{The sigmoid function $\sigma_i\equiv \sigma (\Delta E_i/T)$ [see Eq.~\eqref{eq:boltzmann_machine}] as function of $\Delta E_i$ for $T=0.5,1,2$.}
\label{fig:sigmoid}
\end{figure}

In Eq.~\eqref{eq:boltzmann_machine}, the quantity $\Delta E_i=2 a_i$ is the energy difference between an inactive neuron $i$ and an active one [see Eq.~\eqref{eq:energy_difference_p1_m1}].\footnote{We use the convention employed in Refs.~\refcite{hinton1983optimal,ackley1985learning}. The authors of Ref.~\refcite{hinton1983analyzing}, use the convention that $\Delta E_i = -2 a_i$ and $\sigma_i\equiv \sigma(\Delta E_i/T)=1/(1+\exp(\Delta E_i/T))$.} The function $\sigma(x)=1/\left(1+\exp(-x)\right)$ represents the sigmoid function, and the parameter $T$ serves as an equivalent to temperature.\footnote{In the limit $T\rightarrow 0$, we recover the deterministic update rule \eqref{eq:hopfield_update}.} In Fig.~\ref{fig:sigmoid}, we show $\sigma (\Delta E_i/T)$ as a function of $\Delta E_i$ for $T=0.5,1,2$.  

Examining Eqs.~\eqref{eq:energy_hopfield} and \eqref{eq:boltzmann_machine}, we notice that we are simulating a system akin to the Ising model with Glauber (heat bath) dynamics.\cite{glauber1963time,bottcher2021computational} Because Glauber dynamics satisfy the detailed balance condition, Boltzmann machines will eventually reach thermal equilibrium. The corresponding probabilities $p_{\mathrm{eq}}(X)$ and $p_{\mathrm{eq}}(Y)$ for the ANN to be in states $X$ and $Y$, respectively, will satisfy\footnote{We set the Boltzmann constant $k_B$ to 1.}
\begin{equation}
\frac{p_{\mathrm{eq}}(Y)}{p_{\mathrm{eq}}(X)}=\exp\left( - \frac{E(Y)-E(X)}{T}\right)\,.
\label{eq:bm_boltzmann}
\end{equation}
In other words, the Boltzmann distribution provides the relative probability $p_{\mathrm{eq}}(Y)/p_{\mathrm{eq}}(X)$ associated with the states $X$ and $Y$ of a ``thermalized'' Boltzmann machine. Regardless of the initial configuration, at a given temperature $T$, the stochastic update rule in which neurons are activated with probability $\sigma_i$ always leads to a thermal equilibrium configuration that is solely determined by its energy.
\subsection{Boltzmann machines}
\begin{figure}[t]
\centering
\begin{tikzpicture}[scale=1.3]
    \begin{scope}[rotate=0]
        \foreach \number in {1,...,\nbm}{
            \ifnum\number<4
                \node[neuron, draw=black, fill=red!30] (N-\number) at ({\number*(360/\nbm)}:1.5cm) {$h_\number$};
            \else
                \pgfmathtruncatemacro{\newnumber}{\number-3}
                \node[neuron, draw=black, fill=blue!20] (N-\number) at ({\number*(360/\nbm)}:1.5cm) {$v_\newnumber$};
            \fi
        }

        \foreach \number in {1,...,\nbm}{
            \foreach \y in {1,...,\nbm}{
                \draw[stateTransition] (N-\number) -- (N-\y);
            }
        }
        \end{scope}
\end{tikzpicture}
\caption{Boltzmann machines consist of visible units (blue) and hidden units (red). In the shown example, there are five visible units $\left\{v_i\right\}$ $({i\in\{1,\dots,5\}})$ and three hidden units $\left\{h_j\right\}$ $(j\in\{1,2,3\})$. Similar to Hopfield networks, the network architecture in a Boltzmann machine is complete.}
\label{fig:bm}
\end{figure}
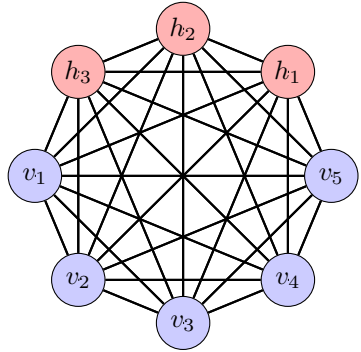

Unlike Hopfield networks, Boltzmann machines have two types of nodes: visible units and hidden units. We denote the corresponding sets of visible and hidden units by $V$ and $H$, respectively. Notice that the set $H$ may be empty. In Fig.~\ref{fig:bm}, we show an example of a Boltzmann machine with five visible units and three hidden units. 

During the training of a Boltzmann machine, the visible units are ``clamped'' to the environment, which means they are set to binary vectors drawn from an empirical distribution. Hidden units may be used to account for constraints involving more than two visible units.

We denote the probability distribution of all configurations of visible units $\{\nu \}$ in a freely running network as $P^\prime \left(\{\nu \}\right)$. Here, ``freely running'' means that no external inputs are fixed (or ``clamped'') to the visible units. We derive the distribution $P^\prime \left(\{\nu \}\right)$ by summing (\ie, marginalizing) over the corresponding joint probability distribution. That is,
\begin{equation}
    P^\prime \left(\{\nu \}\right) = \sum_{\{h\}} P^\prime\left(\{\nu \},\{h\}\right)\,,
\end{equation}
where the summation is performed over all possible configurations of hidden units $\{h \}$.

Our objective is now to devise a method such that $P^\prime \left(\{\nu \}\right)$ converges to the unknown environment (\ie, data) distribution $P\left(\{\nu \}\right)$. To do so, we quantity the disparity between $P^\prime \left(\{\nu \}\right)$ and $P\left(\{\nu \}\right)$ using the Kullback--Leibler (KL) divergence (or relative entropy)
\begin{equation}
    G(P,P^\prime) = \sum_{\{\nu \}} P\left(\{\nu \}\right) \ln\left[\frac{P\left(\{\nu \}\right)}{P^\prime \left(\{\nu \}\right)}\right]\,.
    \label{eq:KL_divergence}
\end{equation}

To minimize the KL divergence $G(P,P^\prime)$, we perform a gradient descent according to
\begin{equation}
    \frac{\partial G}{\partial w_{ij}} = -\frac{1}{T} \left(p_{ij}-p^\prime_{ij}\right)\,,
    \label{eq:grad}
\end{equation}
where $p_{ij}$ represents the probability of both units $i$ and $j$ being active when the environment dictates the states of the visible units, and $p^\prime_{ij}$ is the corresponding probability in a freely running network without any connection to the environment.\cite{hinton1983analyzing,hinton1983optimal,ackley1985learning} We derive Eq.~\eqref{eq:grad} in Sec.~\ref{sec:learning_algorithm}.

Both probabilities $p_{ij}$ and $p_{ij}'$ are measured once the Boltzmann machine has reached thermal equilibrium. Subsequently, the weights $w_{ij}$ of the network are then updated according to
\begin{equation}
\Delta w_{i j} = \epsilon \left(p_{ij}-p^\prime_{ij}\right)\,,
\label{eq:bm_weights}
\end{equation}
where $\epsilon$ is the learning rate. To reach thermal equilibrium, the states of the visible and hidden units are updated using the update probability \eqref{eq:boltzmann_machine}.

In summary, the steps relevant for training a Boltzmann machine are as follows.
\begin{mdframed}
\begin{enumerate}
    \item Clamp the input data (environment distribution) to the visible units.
    \item Update the state of all hidden units according to Eq.~\eqref{eq:boltzmann_machine} until the system reaches thermal equilibrium and then compute $p_{ij}$.
    \item Unclamp the input data from the visible units.
    \item Update the state of all neurons according to Eq.~\eqref{eq:boltzmann_machine} until the system reaches thermal equilibrium and then compute $p_{ij}'$.
    \item Update all weights according to Eq.~\eqref{eq:bm_weights} and return to step 2 or stop if the weight updates are sufficiently small.
\end{enumerate}
\end{mdframed}
After training a Boltzmann machine, we can unclamp the visible units from the environment and generate samples to evaluate their quality. To do this, we use various initial configurations and activate neurons according to Eq.~\eqref{eq:boltzmann_machine} until thermal equilibrium is reached. If the Boltzmann machine was trained successfully, the distribution of states for the unclamped visible units should align with the environment distribution.
\subsection{Restricted Boltzmann machines}
Boltzmann machines are not widely used in general learning tasks. Their impracticality arises from the significant computational burden associated with achieving thermal equilibrium, especially in instances involving large system sizes.
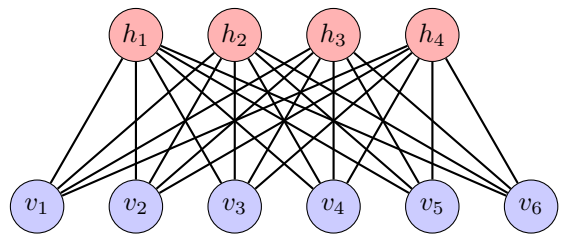
\begin{figure}
\centering
\begin{tikzpicture}[scale=1.3]
    
    \foreach \number in {1,...,4}{
        \node[neuron, draw=black, fill=red!30] (N-\number) at (\number+1, 0) {$h_\number$};
    }
    
    \foreach \number in {1,...,6}{
        \pgfmathtruncatemacro{\newnumber}{\number+4}
        \node[neuron, draw=black, fill=blue!20] (M-\number) at (\number, -1.75) {$v_\number$};
    }
    
    \foreach \x in {1,...,4}{
        \foreach \y in {1,...,6}{
            \draw[stateTransition] (N-\x) -- (M-\y);
        }
    }
\end{tikzpicture}
\caption{Restricted Boltzmann machines consist of a visible layer (blue) and a hidden layer (red). In the shown example, the respective layers comprise six visible units $\left\{v_i\right\}$ $(i\in\{1,\dots,6\})$ and four hidden units $\left\{h_j\right\}$ $(j\in\{1,\dots,4\})$. The network structure of an RBM is bipartite and undirected.}
\label{fig:rbm}
\end{figure}

Restricted Boltzmann machines provide an ANN structure that can be trained more efficiently by omitting connections between the hidden and visible units (see Fig.~\ref{fig:rbm}). Because of these missing intra-layer connections, the network architecture of an RBM is bipartite. 

In RBMs, updates for visible and hidden units are performed alternately. Because there are no connections within each layer, we can update all units within each layer in parallel. Specifically, visible unit $v_i$ is activated with conditional probability
\begin{equation}
p(v_i = 1 | \{h_j\}) = \sigma\Bigl(\sum_{j} w_{ij} h_j + b_i\Bigr)\,,
\end{equation}
where $b_i$ is the bias of visible unit $v_i$ and $\{h_j\}$ is a given configuration of hidden units. We then activate all hidden units based on their conditional probabilities
\begin{equation}
p(h_j = 1 | \{v_i\}) = \sigma\Bigl(\sum_{i } w_{ij}v_{i} + c_j\Bigr)\,,
\end{equation}
where $h_j$ represents hidden unit $j$, $c_j$ is its associated bias, and $\{v_i\}$ denotes a specific configuration of visible units.
This technique of sampling is referred to as ``block Gibbs sampling''.

Training an RBM shares similarities with training a BM. The key difference lies in the need to consider the bipartite network structure in the weight update equation~\eqref{eq:bm_weights}. For an RBM, the weight updates are
\begin{equation}
\Delta w_{i j} = \epsilon (\langle \nu_i h_j \rangle_{\text{data}}-\langle \nu_i h_j \rangle_{\text{model}})\,.
\end{equation}

Instead of sampling configurations to compute $\langle \nu_i h_j \rangle_{\text{data}}$ and $\langle \nu_i h_j \rangle_{\text{model}}$ at thermal equilibrium, we can instead employ a few relaxation steps. This approach is known as ``contrastive divergence''.\cite{hinton2002training,DBLP:conf/aistats/Carreira-Perpinan05} The corresponding weight updates are
\begin{equation}
\Delta w_{i j}^\text{CD} = \epsilon (\langle \nu_i h_j \rangle_{\text{data}}-\langle \nu_i h_j \rangle_{\text{model}}^k)\,.
\end{equation}
The superscript $k$ indicates the number of block Gibbs updates performed. For a more detailed discussion on the contrastive divergence training of RBMs, see Ref.~\refcite{hinton2010practical}.
\begin{figure}
\centering
\includegraphics[width=9cm]{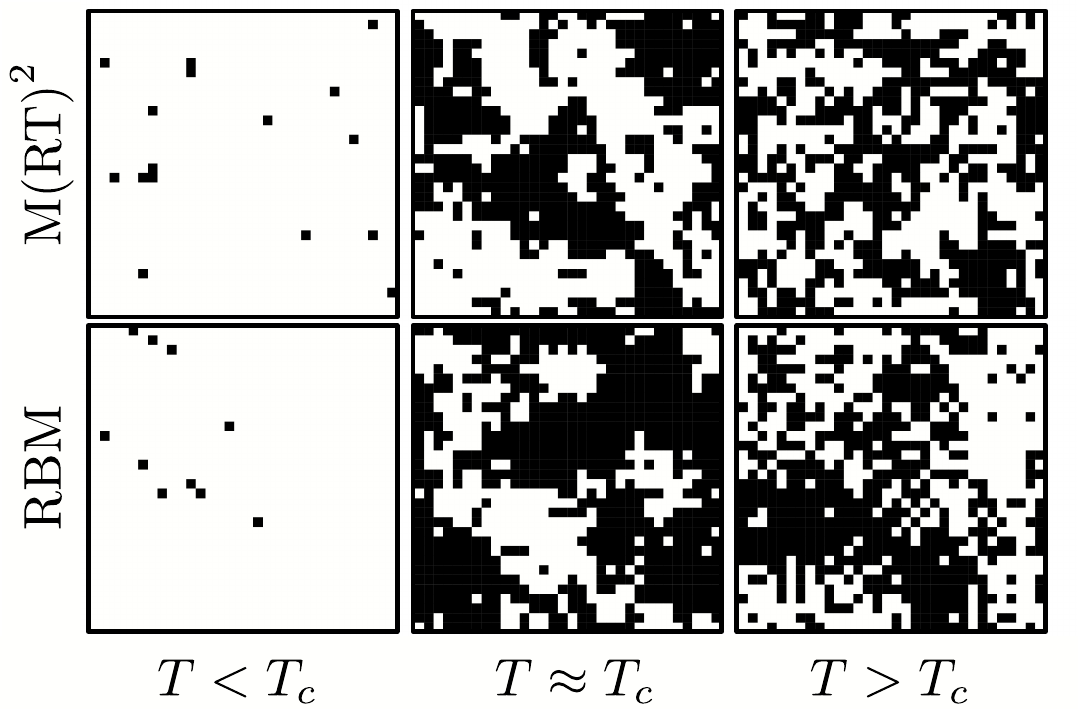}
\caption{Snapshots of $32\times 32$ Ising configurations are shown for $T\in\{1.5, 2.5, 4\}$. These configurations are derived from both M(RT)$^2$ and RBM samples. The quantity $T_c=2/ \ln( 1 + \sqrt{2}) \approx 2.269$ is the critical temperature of the two-dimensional Ising model.}
\label{fig:rbm_ising}
\end{figure}

Restricted Boltzmann machines have been employed in diverse contexts, including dimensionality reduction of datasets\cite{hinton2006reducing}, the study of phase transitions\cite{kim2018smallest, efthymiou2019super, mehta2019high, d2020learning}, and the representation of wave functions\cite{carleo2017solving, torlai2018neural}. In Fig.~\ref{fig:rbm_ising}, we show three snapshots of Ising configurations, generated using both M(RT)$^2$ sampling and RBMs, each comprising $32\times 32$ spins. The RBM was trained using $20\times 10^4$ realizations of Ising configurations sampled at various temperatures.\cite{d2020learning}
\subsection{Derivation of the learning algorithm}
\label{sec:learning_algorithm}
To derive Eq.~\eqref{eq:grad}, we follow the approach outlined in Ref.~\refcite{ackley1985learning}. Notice that the environment distribution $P\left(\{\nu \}\right)$ does not depend $w_{ij}$. Hence, we have
\begin{equation}
\frac{\partial G}{ \partial w_{ij}}=-\sum_{\{\nu \}} \frac{P\left(\{\nu\}\right)}{P^\prime\left(\{\nu\}\right)} \frac{\partial P^\prime\left(\{\nu\}\right)}{\partial w_{i j}}\,.
\label{eq:KL_grad1}
\end{equation}
Next, we wish to compute the gradient $\partial P^\prime(\{\nu\})/\partial w_{ij}$. In a freely running BM, the equilibrium distribution of the visible units follows a Boltzmann distribution. That is,
\begin{equation}
P^\prime (\{\nu\})=\sum_{\{h\}} P^\prime\left(\{\nu\},\{h\}\right)=\frac{\sum_{\{h\}} e^{-E\left({\{\nu, h\}}\right)/T}}{\sum_{\{\nu, h\}} e^{-E\left({\{\nu, h\}}\right)/T}}\,.
\end{equation}
Here, the quantity
\begin{equation}
E\left({\{\nu, h\}}\right)=-\frac{1}{2} \sum_{i,j} w_{i j} x_i^{\{\nu, h\}} x_j^{\{\nu, h\}}
\end{equation}
is the Ising-type energy function in which field (or bias) terms are absorbed in the weights $w_{ij}$ [see Eqs.~\eqref{eq:energy_hopfield} and \eqref{eq:energy_difference2}]. For a BM in state ${\{\nu, h\}}$, we denote the state of neuron $i$ by $x_i^{\{\nu, h\}}$. Using
\begin{equation}
\frac{\partial e^{-E\left({\{\nu, h\}}\right)/T}}{\partial w_{ij}}=\frac{1}{T} x_i^{\{\nu, h\}} x_j^{\{\nu, h\}} e^{-E\left({\{\nu, h\}}\right)/T}
\end{equation}
yields
\begin{align}
\begin{split}
\frac{\partial P^\prime \left(\{\nu\}\right)}{ \partial w_{ij}}&=\frac{\frac{1}{T}\sum_{\{h\}} x_i^{\{\nu, h\}} x_j^{\{\nu, h\}} e^{-E\left({\{\nu, h\}}\right)/T}}{\sum_{\{\nu, h\}} e^{-E\left({\{\nu, h\}}\right)/T}}\\
&-\frac{\sum_{\{h\}} e^{-E\left({\{\nu, h\}}\right)/T} \frac{1}{T} \sum_{\{\nu, h\}} x_i^{\{\nu, h\}} x_j^{\{\nu, h\}} e^{-E\left({\{\nu, h\}}\right)/T}}{\left(\sum_{\{\nu, h\}} e^{-E\left({\{\nu, h\}}\right)/T}\right)^2}\\
&=\frac{1}{T} \left[ \sum_{\{h\}}x_i^{\{\nu, h\}} x_j^{\{\nu, h\}} P^\prime\left(\{\nu\},\{h\}\right)\right. \\
&\hspace{1cm}\left.-P^\prime(\{\nu\})\sum_{\{\nu, h\}} x_i^{\{\nu, h\}} x_j^{\{\nu, h\}}P^\prime\left(\{\nu\},\{h\}\right) \right]\,.
\end{split}
\end{align}
We will now substitute this result into Eq.~\eqref{eq:KL_grad1} to obtain
\begin{align}
\begin{split}
\frac{\partial G}{ \partial w_{ij}}&=-\sum_{\{\nu\}} \frac{P\left(\{\nu\}\right)}{P^\prime\left(\{\nu\}\right)} \frac{1}{T} \left[ \sum_{\{h\}} x_i^{\{\nu, h\}} x_j^{\{\nu, h\}} P^\prime\left(\{\nu\}, \{h\}\right)\right. \\
&\hspace{3.2cm}\left.-P^\prime\left(\{\nu\}\right)\sum_{\{\nu, h\}} x_i^{\{\nu, h\}} x_j^{\{\nu, h\}} P^\prime\left(\{\nu\}, \{h\}\right)\right]\\
&=-\frac{1}{T} \left[ \sum_{\{\nu, h\}} x_i^{\{\nu, h\}} x_j^{\{\nu, h\}} P\left(\{\nu\}, \{h\}\right)-\sum_{\{\nu, h\}} x_i^{\{\nu, h\}} x_j^{\{\nu, h\}} P^\prime\left(\{\nu\}, \{h\}\right)\right]\,,
\end{split}
\end{align}
where we used that $\sum_{\{\nu \}} P\left(\{\nu\}\right)=1$ and 
$P\left(\{\nu\}\right)/P^\prime\left(\{\nu\}\right) P^\prime\left(\{\nu\},\{h\}\right) = P\left(\{\nu\},\{h\}\right)\,.
$
The latter equation follows from the definition of joint probability distributions
\begin{equation}
P\left(\{\nu\},\{h\}\right)=P\left(\{h\}|\{\nu\}\right)P\left(\{\nu\}\right)
\end{equation}
and
\begin{equation}
    P^\prime\left(\{\nu\},\{h\}\right)=P^\prime\left(\{h\}|\{\nu\}\right)P^\prime\left(\{\nu\}\right)\,,
\end{equation}
with $P\left(\{h\}|\{\nu\}\right)=P^\prime\left(\{h\}|\{\nu\}\right)$. The conditional distributions $P(\{h\}|\{\nu\})$ and $P^\prime(\{h\}|\{\nu\})$ are identical, as the probability of observing a certain hidden state given a visible one is independent of the origin of the visible state in an equilibrated system. In other words, for the conditional distributions in an equilibrated system, it is irrelevant whether the visible state is provided by the environment or generated by a freely running machine. Defining
\begin{equation}
p_{ij} \coloneqq \sum_{\{\nu, h\}} x_i^{\{\nu, h\}} x_j^{\{\nu, h\}} P\left(\{\nu\},\{h\}\right)
\end{equation}
and
\begin{equation}
    p_{ij}^\prime \coloneqq \sum_{\{\nu, h\}} x_i^{\{\nu, h\}} x_j^{\{\nu, h\}} P^\prime\left(\{\nu\},\{h\}\right)\,,
\end{equation}
we finally obtain Eq.~\eqref{eq:grad}.
\section{Loss landscapes of artificial neural networks}
\label{sec:loss_landscapes}
Training an ANN involves minimizing a given loss function such as the KL divergence [see Eq.~\eqref{eq:KL_divergence}]. The loss landscapes of ANNs are affected by several factors, including structural properties \cite{hornik1989multilayer,hornik1991approximation,DBLP:conf/iclr/ParkYLS21} and a range of implementation attributes \cite{hardt2016train,wilson2017marginal,choi2020on}. 
Knowledge of their precise effects on learning performance, however, remains incomplete.

One path to a better understanding of the relationships between ANN structure, implementation attributes, and learning performance is through a more in-depth analysis of the geometric properties of loss landscapes. For instance, Keskar \etal~\cite{DBLP:conf/iclr/KeskarMNST17} analyze the local curvature around candidate minimizers via the spectrum of the underlying Hessian to characterize the flatness and sharpness of loss minima, and Dinh \etal~\cite{DBLP:conf/icml/DinhPBB17} demonstrate that reparameterizations can render flat minima sharp without affecting generalization properties. Even so, one challenge to the study of geometric properties of loss landscapes is high dimensionality.   To meet that challenge, some propose to visualize the curvature around a given point by projecting curvature properties of high-dimensional loss functions to a lower-dimensional (and often random) projection of two or three dimensions~\cite{DBLP:journals/corr/GoodfellowV14,DBLP:conf/nips/Li0TSG18,DBLP:conf/emnlp/HaoDWX19,wu2020adversarial,DBLP:conf/ida/HoroiHRLWK22}. Horoi \etal~\cite{DBLP:conf/ida/HoroiHRLWK22}, building on this approach, pursue improvements in learning by dynamically sampling points in projected low-loss regions surrounding local minima during training.

However, visualizing high-dimensional loss landscape curvature relies on accurate projections of curvature properties to lower dimensions. Unfortunately, random projections do not preserve curvature information, thus do not afford accurate low-dimensional representations of curvature information. This argument is given in Sec.~\ref{sec:projection_curvature} and illustrated with a simulation example in Sec.~\ref{sec:examples_extracting_curvature_info}.  Principal curvatures in a low-dimensional projection {\em are} nevertheless given by functions of weighted ensemble means of the Hessian elements in the original, high-dimensional space, a result also established in Sec.~\ref{sec:projection_curvature}. Instead of using random projections to visualize loss functions, we propose to analyze projections along {\em dominant Hessian directions} associated with the largest-magnitude positive and negative principal curvatures.  

\subsection{Principal curvature in random projections}
\label{sec:projection_curvature}
To describe the connection between the principal curvature of a loss function $L(\theta)$ with ANN parameters $\theta\in\mathbb{R}^N$ and that associated with a lower-dimensional, random projection, we provide in Sec.~\ref{sec:overview} an overview of concepts from differential geometry~\cite{lee2006riemannian,berger2012differential,kuhnel2015differential} that are useful to mathematically describe curvature in high-dimensional spaces. In Secs.~\ref{sec:random_projections} and \ref{sec:principal_curvature}, we  show the relationship between principal curvature and random projections. Finally, in Sec.~\ref{sec:hessian_trace_estimates}, we highlight a relationship between measures of curvature presented in Sec.~\ref{sec:principal_curvature} and Hessian trace estimates, which affords an alternative to Hutchinson's method for computing unbiased Hessian trace estimates.
\subsubsection{Differential and information geometry concepts}
\label{sec:overview}
In differential geometry, the principal curvatures are the eigenvalues of the shape operator (or Weingarten map\footnote{Some authors distinguish between the shape operator and the Weingarten map depending on if the change of the underlying tangent vector is described in the original manifold or in Euclidean space (see, \eg, chapter 3 in \cite{crane2018discrete}).}), a linear endomorphism defined on the tangent space $T_p$ of $L$ at a point $p$. For the original high-dimensional space, we have $(\theta,L(\theta))\subseteq \mathbb{R}^{N+1}$ and there are $N$ principal curvatures $\kappa_1^\theta\geq \kappa_2^\theta \geq \dots \geq \kappa_N^\theta$. At a non-degenerate critical point $\theta^*$ where the gradient $\nabla_\theta L$ vanishes, the matrix of the shape parameter is given by the Hessian $H_\theta$ with elements $(H_\theta)_{ij}=\partial^2 L/(\partial \theta_i \partial \theta_j)$ ($i,j\in\{1,\dots, N\}$).\footnote{A critical point is degenerate if the Hessian $H_{\theta}$ at this point is singular (\ie, $\mathrm{det}(H_{\theta})=0$). At degenerate critical points, one cannot use the eigenvalues of $H_{\theta}$ to determine if the critical point is a minimum (positive definite $H_{\theta}$) or a maximum (negative definite $H_{\theta}$). Geometrically, at a degenerate critical point, a quadratic approximation fails to capture the local behavior of the function that one wishes to study.} Some works refer to the Hessian as the ``curvature matrix''~\cite{DBLP:journals/jmlr/Martens20} or use it to characterize the curvature properties of $L(\theta)$~\cite{DBLP:conf/nips/Li0TSG18}. In the vicinity of a non-degenerate critical point $\theta^*$, the eigenvalues of the Hessian $H_\theta$ are the principal curvatures and describe the loss function in the eigenbasis of $H_\theta$ according to
\begin{equation}
L(\theta^*+\Delta \theta)=L(\theta^*)+\frac{1}{2}\sum_{i=1}^N \kappa_i^{\theta}\Delta \theta_i^2\,.
\end{equation}
The Morse lemma states that, if a critical point $\theta^*$ of $L(\theta)$ is non-degenerate, then there exists a chart $(\tilde{\theta}_1,\dots,\tilde{\theta}_N)$ in a neighborhood of $\theta^*$ such that
\begin{equation}
\label{lem:Morse}
L(\tilde{\theta}) = -\tilde{\theta}^2_1 - \cdots - \tilde{\theta}^2_i + \tilde{\theta}^2_{i +1} + \cdots + \tilde{\theta}^2_N + L(\theta^*)\,, 
\end{equation}
where $\tilde{\theta}_i(\theta)=0$ for $i\in\{1,\dots,N\}$. The loss function $L(\tilde{\theta})$ in Eq.~\eqref{lem:Morse} is decreasing along $i$ directions and increasing along the remaining $i+1$ to $N$ directions. Further, the index $i$ of a critical point $\theta^*$ is the number of negative eigenvalues of the Hessian $H_\theta$ at that point.

In the standard basis, the Hessian is
\begin{equation}
H_{\theta}\coloneqq \nabla_\theta \nabla_\theta L(\theta)=\begin{pmatrix}
\frac{\partial^2 L}{\partial \theta_1^2} & \cdots & \frac{\partial^2 L}{\partial \theta_1 \partial \theta_N} \\
\vdots  & \ddots & \vdots \\
\frac{\partial^2 L}{\partial \theta_N \partial \theta_1} & \cdots & \frac{\partial^2 L}{\partial \theta_N^2}
\end{pmatrix}\,.
\end{equation}

\subsubsection{Random projections}
\label{sec:random_projections}
To graphically explore an $N$-dimensional loss function $L$ around a critical point $\theta^*$, one may wish to work in a lower-dimensional representation. For example, a two-dimensional projection of $L$ around $\theta^*$ is provided by
\begin{equation}
L(\theta^*+\alpha\eta+\beta\delta)\,,
\label{eq:app_loss_alpha_beta}
\end{equation}
where the parameters $\alpha,\beta\in \mathbb{R}$ scale the directions $\eta,\delta\in\mathbb{R}^N$. The corresponding graph representation is $(\alpha,\beta,L(\alpha,\beta))\subseteq\mathbb{R}^{3}$. 

In high-dimensional spaces, there exist vastly many more almost-orthogonal than orthogonal directions. In fact, if the dimension of our space is large enough, with high probability, random vectors will be sufficiently close to orthogonal \cite{HechtNielsen94}. Following this result, many related works~\cite{DBLP:conf/nips/Li0TSG18,wu2020adversarial,DBLP:conf/ida/HoroiHRLWK22} use random Gaussian directions with independent and identically distributed vector elements $\eta_i, \delta_i \sim\mathcal{N}(0,1)$ ($i\in\{1,\dots,N\}$).


The scalar product of random Gaussian vectors $\eta,\delta$ is a sum of the difference between two chi-squared distributed random variables because
\begin{equation}
\sum_{i=1}^N \eta_i \delta_i=\sum_{i=1}^N \frac{1}{4}(\eta_i+\delta_i)^2-\frac{1}{4}(\eta_i-\delta_i)^2=\sum_{i=1}^N \frac{1}{2} X_i^2-\frac{1}{2} Y_i^2\,, 
\label{eq:scalar_product_eta_delta}
\end{equation}
where $X_i,Y_i\sim \mathcal{N}(0,1)$.

Notice that $\eta,\delta$ are almost orthogonal, which can be proved using a concentration inequality for chi-squared distributed random variables.\footnote{For further details, see Ref.~\refcite{bottcher2024visualizing}.}

\subsubsection{Principal curvature}
\label{sec:principal_curvature}

With the form of random Gaussian projections in hand, we now analyze the principal curvatures in both the original and lower-dimensional spaces. The Hessian associated with the two-dimensional loss projection \eqref{eq:app_loss_alpha_beta} is
\begin{align}
\begin{split}
H_{\alpha,\beta}&=\begin{pmatrix}
\frac{\partial^2 L}{\partial \alpha^2} & \frac{\partial^2 L}{\partial \alpha \partial \beta} \\
 \frac{\partial^2 L}{\partial \beta \partial \alpha}  & \frac{\partial^2 L}{\partial \beta^2} \\
\end{pmatrix}\\
&=
\begin{pmatrix}
\sum_{i,j} \eta_i \eta_j \frac{\partial^2 L}{\partial \theta_i \theta_j} & \sum_{i,j} \eta_i \delta_j \frac{\partial^2 L}{\partial \theta_i \theta_j} \\
\sum_{i,j} \eta_i \delta_j \frac{\partial^2 L}{\partial \theta_i \theta_j}  & \sum_{i,j} \delta_i \delta_j \frac{\partial^2 L}{\partial \theta_i \theta_j} \\
\end{pmatrix}\\
&=\begin{pmatrix}
(H_\theta)_{ij} \eta^i \eta^j & (H_\theta)_{ij} \eta^i \delta^j\\
(H_\theta)_{ij} \eta^i \delta^j & (H_\theta)_{ij} \delta^i \delta^j\\
\end{pmatrix}\,,
\end{split}
\label{eq:hessian_2d}
\end{align}
where we use Einstein notation in the last equality.


Because the elements of $\delta,\eta$ are distributed according to a standard normal distribution, the second derivatives of the loss function $L$ in Eq.~\eqref{eq:hessian_2d} have prefactors that are products of standard normal variables and, hence, can be expressed as sums of chi-squared distributed random variables as in Eq.~\eqref{eq:scalar_product_eta_delta}. To summarize, elements of $H_{\alpha,\beta}$ are sums of second derivatives of $L$ in the original space weighted with chi-squared distributed prefactors. 

The principal curvatures $\kappa_{\pm}^{\alpha,\beta}$ (\ie, the eigenvalues of $H_{\alpha,\beta}$) are 
\begin{equation}
\kappa_{\pm}^{\alpha,\beta}=\frac{1}{2}\left(A+C\pm \sqrt{4 B^2+(A-C)^2}\right)\,,
\label{eq:kapp_alpha_beta}
\end{equation}
where $A=(H_\theta)_{ij}\eta^i \eta^j $, $B=(H_\theta)_{ij} \eta^i \delta^j$, and $C=(H_\theta)_{ij} \delta^i \delta^j $. To the best of our knowledge, a closed, analytic expression for the distribution of the quantities $A,B,C$ is not yet known~\cite{davies1980algorithm,laurent2000adaptive,bausch2013efficient,chen2019numerical}.

Returning to principal curvature, since $\sum_{i,j} a_{ij} \eta^{i} \eta^{j}=\sum_{i} a_{ii} \eta^{i} \eta^{i}+\sum_{i\neq j} a_{ij} \eta^i \eta^j$ ($a_{ij}\in\mathbb{R}$), 
we find that $\mathds{E}[A]=\mathds{E}[C]={(H_\theta)^i}_i$ 
and $\mathds{E}[B]=0$ where ${(H_\theta)^i}_i\equiv \mathrm{tr}(H_\theta)=\sum_{i=1}^N \kappa_i^\theta$.\footnote{The expected values of the quantities $A$, $B$, and $C$ correspond to ensemble means \eqref{eq:ensemble_average} in the limit $S\rightarrow\infty$, where $S$ is the number of independent realizations of the underlying random variable.} To show that the expected values of $a_{ij}\eta^i\eta^j$ ($i\neq j$) or $a_{ij}\eta^i\delta^j$ vanish, one can either invoke independence of $\eta^i,\eta^j$ ($i\neq j$) and $\eta^i,\delta^j$ or transform both products into corresponding differences of two chi-squared random variables with the same mean [see Eq.~\eqref{eq:scalar_product_eta_delta}].

Hence, the expected, dimension-reduced Hessian \eqref{eq:hessian_2d} is
\begin{equation}
\mathds{E}[H_{\alpha,\beta}]=\begin{pmatrix} {(H_\theta)^i}_i& 0\\
0 & {(H_\theta)^i}_i\\
\end{pmatrix}\,.
\label{eq:expected_hessian}
\end{equation}
The corresponding eigenvalue (or principal curvature) $\bar{\kappa}^{\alpha,\beta}$ is therefore given by the sum over all principal curvatures in the original space (\ie, $\bar{\kappa}^{\alpha,\beta}=\sum_{i=1}^N \kappa_i^\theta$). Hence, the value of the principal curvature $\bar{\kappa}^{\alpha,\beta}$ in the expected dimension-reduced space will be either positive (if the positive principal curvatures in the original space dominate), negative (if the negative principal curvatures in the original space dominate), or close to zero (if positive and negative principal curvatures in the original space cancel out each other). As a result, saddle points will not appear as such in the expected random projection.

In addition to the connection between $\bar{\kappa}^{\alpha,\beta}$ and the principal curvatures $\kappa_i^\theta$, we now provide an overview of additional mathematical relations between different curvature measures that are useful to quantify curvature properties of high-dimensional loss functions and their two-dimensional random projections.

\begin{table}[t]
\tbl{Quantities to characterize curvature.}{
\label{tab:curvature_measures}
\begin{tabular}{l l}
\toprule
{\sc Symbol} &{\sc Definition} \\
\midrule
$H_\theta\in\mathbb{R}^{N\times N}$  & 
Hessian in original loss space\\[2pt]
$\kappa_{i}^{\theta}\in\mathbb{R}$  & principal curvatures in original loss space \\
 		& with $i\in\{1,\dots,N\}$ (\ie, the eigenvalues of $H_{\theta}$)\\[2pt]
$H_{\alpha,\beta}\in\mathbb{R}^{2\times 2}$ &  Hessian in a two-dimensional projection\\
		& of an $N$-dimensional loss function\\[2pt]
$\kappa_{\pm}^{\alpha,\beta}\in\mathbb{R}$ & principal curvatures in a two-dimensional\\
		& loss projection (\ie, the eigenvalues of $H_{\alpha,\beta}$)\\[2pt]
$\bar{\kappa}^{\alpha,\beta}\in\mathbb{R}$  & principal curvature in expected, two-dimensional\\
		&  loss projection (\ie, the eigenvalues of $\mathbb{E}[H_{\alpha,\beta}]$)\\[2pt]
$H\in\mathbb{R}$ & mean curvature~(\ie, $\sum_{i=1}^N \kappa_i^\theta/N=\bar{\kappa}^{\alpha,\beta}/N$)\\
\bottomrule
\end{tabular}}
\end{table}

Invoking Eq.~\eqref{eq:kapp_alpha_beta}, we can relate $\bar{\kappa}^{\alpha,\beta}$ to $\mathrm{tr}(H_\theta)$ and $\kappa_{\pm}^{\alpha,\beta}$. Because $\kappa_{+}^{\alpha,\beta}+\kappa_{-}^{\alpha,\beta}=A+C$, we have
\begin{equation}
\mathrm{tr}(H_\theta)=\bar{\kappa}^{\alpha,\beta}=\sum_{i=1}^N \kappa_i^\theta=\frac{1}{2}\left(\mathbb{E}[\kappa_{-}^{\alpha,\beta}]+\mathbb{E}[\kappa_{+}^{\alpha,\beta}]\right)\,.
\label{eq:curvature_relations}
\end{equation}
The mean curvature $H$ in the original space is related to $\bar{\kappa}^{\alpha,\beta}$ via
\begin{equation}
H=\frac{1}{N}\bar{\kappa}^{\alpha,\beta}=\frac{1}{N}\sum_{i=1}^N \kappa_i^\theta\,.
\label{eq:mean_curvature}
\end{equation}
We summarize the definitions of the employed Hessians and curvature measures in Tab.~\ref{tab:curvature_measures}.

The appeal of random projections is that pairwise distances between points in a high-dimensional space can be nearly preserved by a lower-dimensional linear embedding, affording a low-dimensional representation of mean and variance information with minimal distortion \cite{Matousek:2008}. The relationship between random Gaussian directions and principal curvature is less straightforward. Our results show that the principal curvatures $\kappa_{\pm}^{\alpha,\beta}$ in a two-dimensional loss projection are weighted averages of the Hessian elements $(H_\theta)_{ij}$ in the original space, not weighted averages of the principal curvatures  $\kappa_i^\theta$ as claimed by Ref.~\refcite{DBLP:conf/nips/Li0TSG18}. Similar arguments apply to projections with dimension larger than 2. 
\subsubsection{Hessian trace estimates}
\label{sec:hessian_trace_estimates}
Finally, we point to a connection between curvature measures and Hessian trace estimates. A common way of estimating $\mathrm{tr}(H_\theta)$ without explicitly computing all eigenvalues of $H_\theta$ is based on Hutchinson's method~\cite{hutchinson1989stochastic} and random numerical linear algebra~\cite{bai1996some,avron2011randomized}. The basic idea behind this approach is to (i) use a random vector $z\in\mathbb{R}^N$ with elements $z_i$ that are distributed according to a distribution function with zero mean and unit variance (\eg, a Rademacher distribution with $\Pr\left(z_i=\pm 1\right)=1/2$), and (ii) compute $z^\top H_\theta z$, an unbiased estimator of $\mathrm{tr}(H_\theta)$. That is, 
\begin{equation}
\mathrm{tr}(H_\theta)=\mathbb{E}[z^\top H_\theta z]\,.
\label{eq:hutchinson}
\end{equation}
Recall that Eq.~\eqref{eq:curvature_relations} shows that the principal curvature of the expected random loss projection, $\bar{\kappa}^{\alpha,\beta}$, is equal to $\mathrm{tr}(H_\theta)$. Instead of estimating $\mathrm{tr}(H_\theta)$ using Hutchinson's method \eqref{eq:hutchinson}, an alternative Hutchinson-type estimate of this quantity is provided by the mean of the expected values of $\kappa_{-}^{\alpha,\beta}$ and $\kappa_{+}^{\alpha,\beta}$ [see Eq.~\eqref{eq:curvature_relations}].

\subsection{Extracting curvature information}
\label{sec:examples_extracting_curvature_info}
We now study two examples that will help build intuitions. In the first example presented in Sec.~\ref{sec:equal_number_curvature}, we study a critical point $\theta^*$ of an $N$-dimensional loss function $L(\theta)$ for which (i) all principal curvatures have the same magnitude and (ii) the number of positive curvature directions is equal to the number of negative curvature directions. In this example, saddles can be correctly detected by ensemble means but success depends on the averaging process used. In the second example presented in Sec.~\ref{sec:unequal_number_curvature}, we use a loss function associated with an unequal number of negative and positive curvature directions. For the different curvature measures derived in Sec.~\ref{sec:principal_curvature}, we find that random projections cannot identify the underlying saddle point. In Sec.~\ref{sec:hessian_trace}, we will use the two example loss functions to discuss how curvature-based Hessian trace estimates relate to those obtained with the original Hutchinson's method.
\subsubsection{Equal number of curvature directions}
\label{sec:equal_number_curvature}
The loss function of our first example is
\begin{equation}
L(\theta)=\frac{1}{2}\theta_{2n+1}\left(\sum_{i=1}^n\theta_i^2-\theta_{i+n}^2\right)\,,\quad n\in\mathbb{Z}_{+}\,,
\label{eq:loss_symmetric}
\end{equation}
where we set $N=2n+1$. A critical point $\theta^*$ of the loss function \eqref{eq:loss_symmetric} satisfies
\begin{equation}
    (\nabla_\theta L)(\theta^*)=
    \left(
    \begin{array}{c}
    \theta^*_1 \theta^*_{2n+1}\\
    \vdots\\
    \theta^*_{n}\theta^*_{2n+1}\\
    -\theta^*_{n+1}\theta^*_{2n+1}\\
    \vdots\\
    -\theta^*_{2n}\theta^*_{2n+1}\\
    \frac{1}{2}\left(\sum_{i=1}^n{\theta_{i}^{{*\textsuperscript{2}}}}-{\theta_{i+n}^{{*\textsuperscript{2}}}}\right)
    \end{array}
    \right)=0\,.
\end{equation}
The Hessian at the critical point $\theta^*=(\theta^*_1,\dots,\theta^*_{2n},\theta^*_{2n+1})$ $=(0,\dots,0,1)$ is 
\begin{equation}
H_\theta=\mathrm{diag}(\underbrace{1,\dots,1}_{n~\mathrm{times}},\underbrace{-1,\dots,-1}_{n~\mathrm{times}},0)\,.
\label{eq:emp_hessian_1}
\end{equation}
\begin{figure}
    \centering
    \includegraphics[width=.475\textwidth]{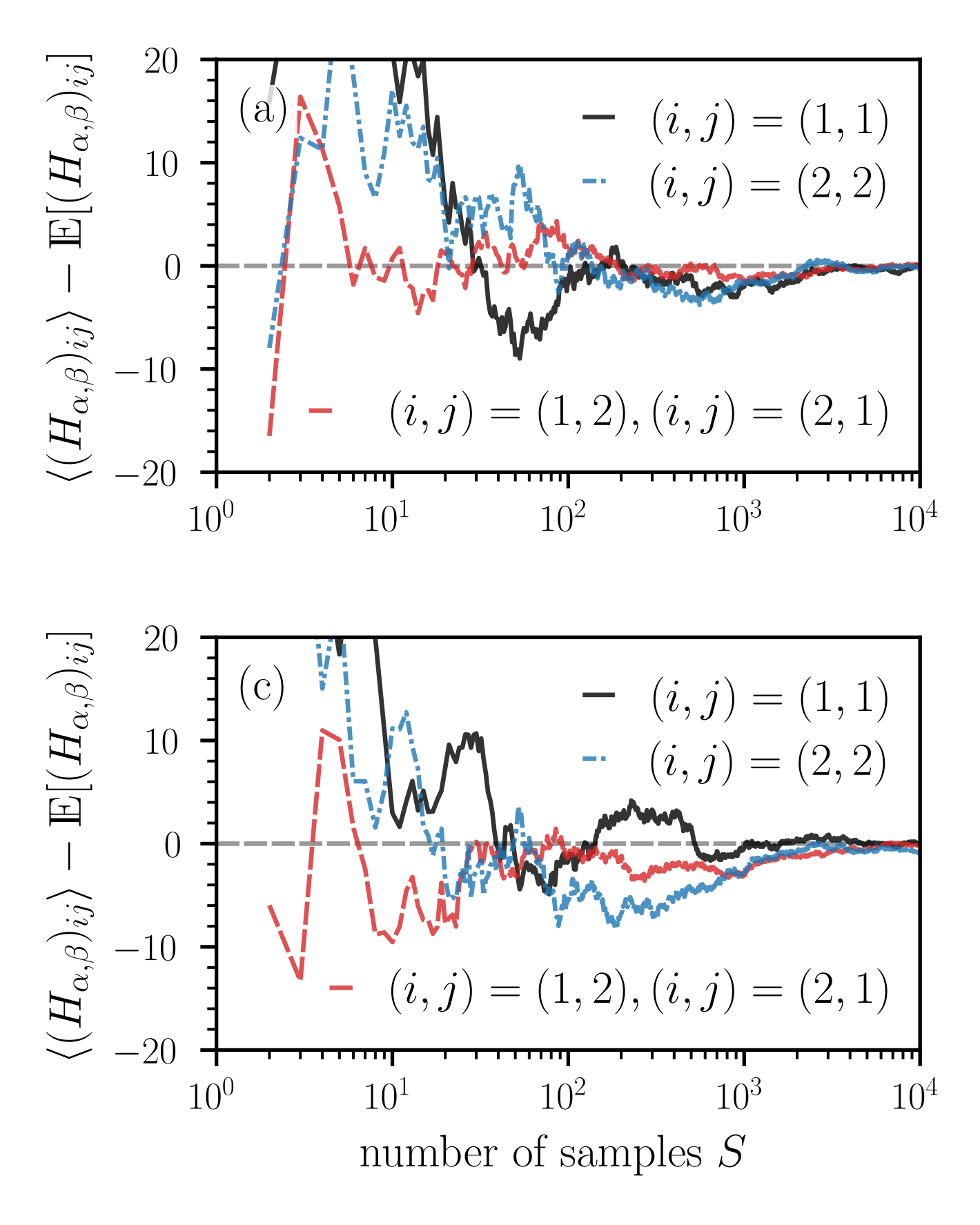}
    \includegraphics[width=.475\textwidth]{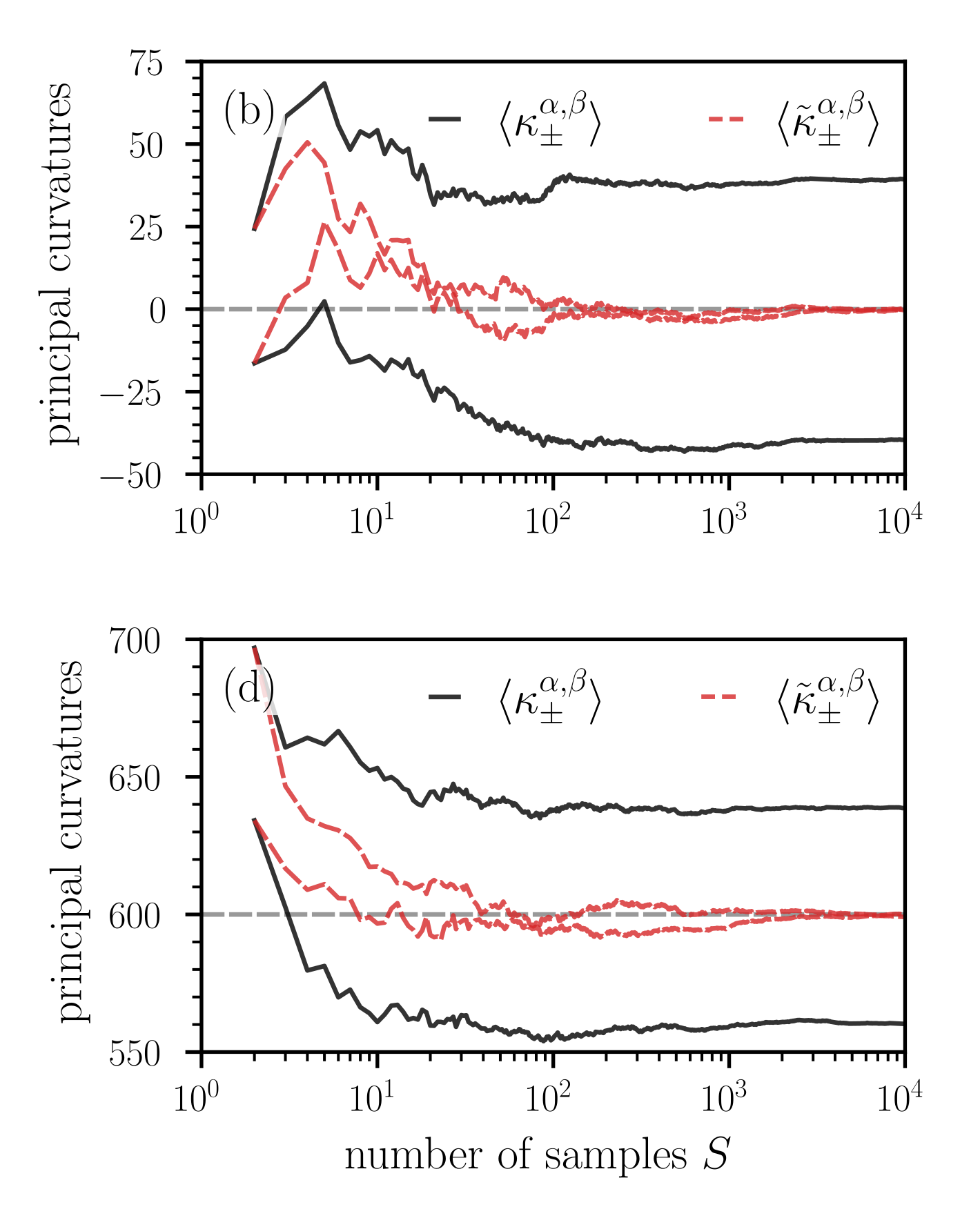}
    \caption{Convergence of the ensemble mean \eqref{eq:ensemble_average} of Hessian elements and curvatures measures as a function of the number of random projections $S$. (a,c) The deviation of the ensemble means $\langle (H_{\alpha,\beta})_{ij}\rangle$ (${i,j\in\{1,2\}}$) of Hessian elements from the corresponding expected values as a function of $S$. Notice that the expected value of the diagonal elements $(H_{\alpha,\beta})_{11}$ and $(H_{\alpha,\beta})_{22}$ is equal to $\bar{\kappa}^{\alpha,\beta}$ (\ie, to the sum of principal curvatures in the original space) [see Eqs.~\eqref{eq:expected_hessian} and \eqref{eq:curvature_relations}]. A relatively large number of random projections between $10^3$ and $10^4$ is required to keep the deviations at values smaller than about 2--4. (b,d) The ensemble means $\langle \kappa^{\alpha,\beta}_{\pm}\rangle$ [see Eq.~\eqref{eq:kapp_alpha_beta}] and $\langle \tilde{\kappa}^{\alpha,\beta}_{\pm}\rangle$ [see Eq.~\eqref{eq:kappa_tilde_sample}] as a function of $S$. Dashed grey lines represent $\bar{\kappa}^{\alpha,\beta}=\mathrm{tr}(H_\theta)$. In panels (a,b) and (c,d), the $N$-dimensional loss functions are given by Eqs.~\eqref{eq:loss_symmetric} and \eqref{eq:loss_asymmetric}, respectively. We evaluate the corresponding Hessians \eqref{eq:emp_hessian_1} and \eqref{eq:emp_hessian_2} at the saddle point $\theta^*=(\theta^*_1,\dots,\theta^*_{2n},\theta^*_{2n+1})=(0,\dots,0,1)$. In both loss functions, we set $n=500$ and in loss function~\eqref{eq:loss_asymmetric} we set $\tilde{n}=800$.}
    \label{fig:hessians_random}
\end{figure}
Because $H_\theta$ has positive and negative eigenvalues, the critical point is a saddle. The corresponding principal curvatures are $\kappa_i^\theta\in\{-1,1\}$ ($i\in\{1,\dots,N-1\}$) and $\kappa^\theta_N=0$. In this example, the mean curvature $H$, as defined in Eq.~\eqref{eq:mean_curvature}, is equal to 0. According to Eq.~\eqref{eq:expected_hessian}, the principal curvature, $\bar{\kappa}^{\alpha,\beta}$, associated with the expected, dimension-reduced Hessian $H_{\alpha,\beta}$ is also equal to 0, erroneously indicating an apparently flat loss landscape if one would use $\bar{\kappa}^{\alpha,\beta}$ as the main measure of curvature. To compare the convergence of different curvature measures as a function of the number of loss projections $S$, we will now study the ensemble mean
\begin{equation}
\langle X \rangle=\frac{1}{S}\sum_{k=1}^S X^{(k)}
\label{eq:ensemble_average}
\end{equation}
of different quantities of interest $X$ such as Hessian elements and principal curvature measures in dimension-reduced space. Here, $X^{(k)}$ is the $k$-th realization (or sample) of $X$.
\begin{figure}
    \centering
    \includegraphics[width=.475\textwidth]{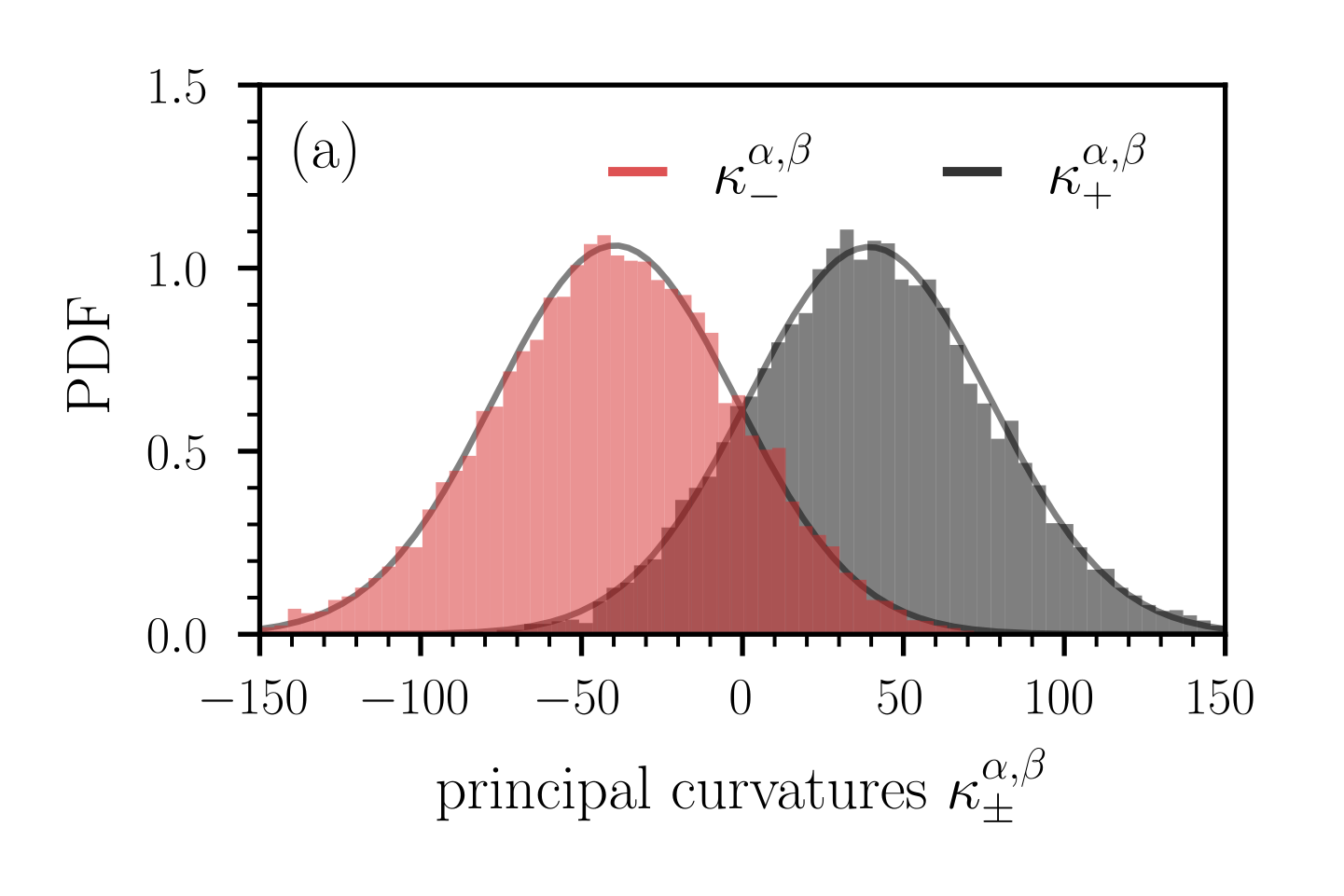}
     \includegraphics[width=.475\textwidth]{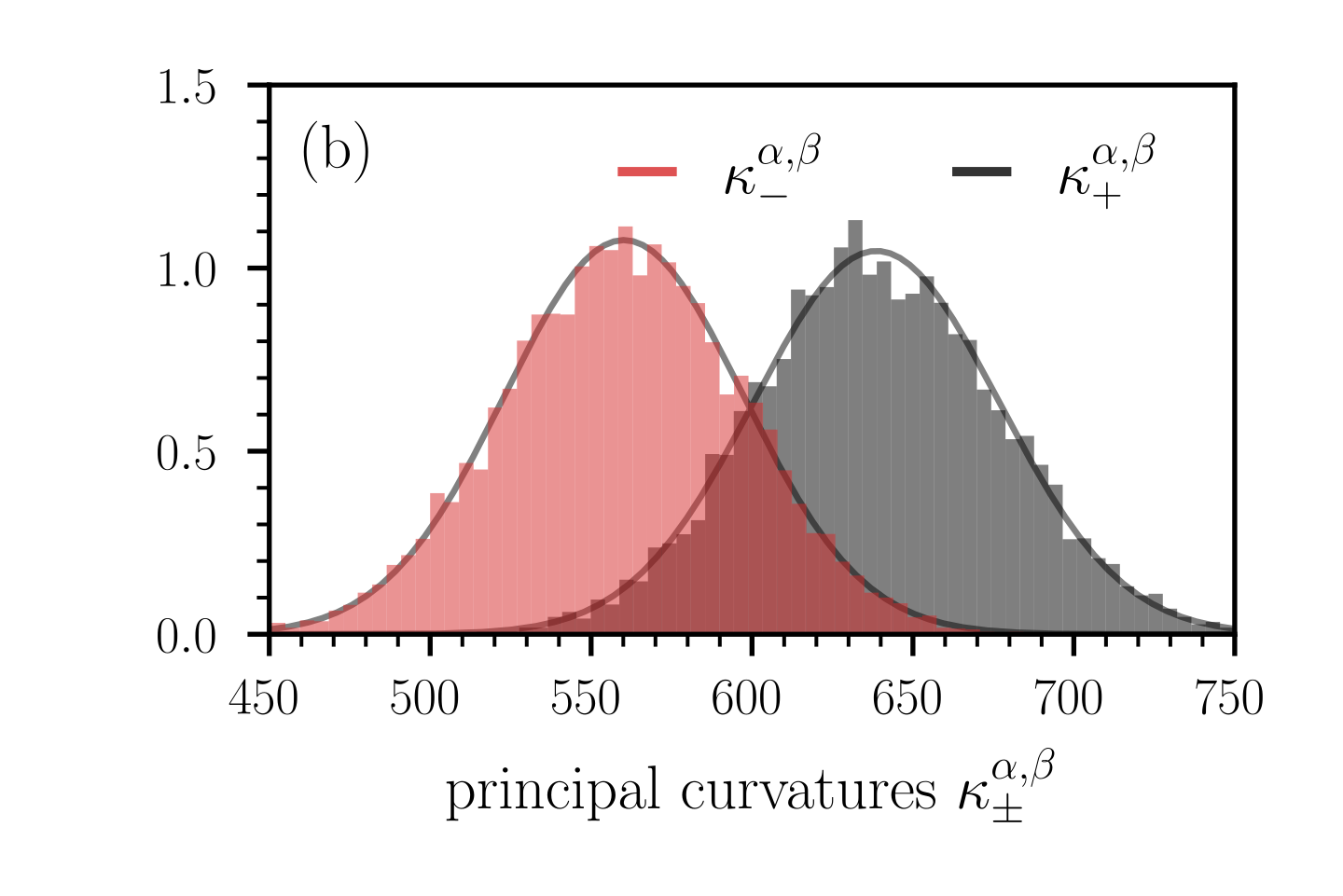}
    \caption{Distribution of principal curvatures $\kappa_{-}^{\alpha,\beta}$ (red bars) and $\kappa_{+}^{\alpha,\beta}$ (black bars). In panels (a) and (b), the loss functions are given by Eqs.~\eqref{eq:loss_symmetric} and \eqref{eq:loss_asymmetric}, respectively. We evaluate the corresponding Hessians \eqref{eq:emp_hessian_1} and \eqref{eq:emp_hessian_2} at the saddle point $\theta^*=(\theta^*_1,\dots,\theta^*_{2n},\theta^*_{2n+1})=(0,\dots,0,1)$. In both loss functions, we set $n=500$ and in loss function~\eqref{eq:loss_asymmetric} we set $\tilde{n}=800$. In panel (a), the probability $\Pr(\kappa_+^{\alpha,\beta}\kappa_-^{\alpha,\beta}>0)$ that the critical point in the lower-dimensional, random projection does not appear as a saddle is about 0.3, and it is 1 in panel (b). 
    Histograms are based on 10,000 random projections used to compute $\kappa_{\pm}^{\alpha,\beta}$. Solid grey lines are Gaussian approximations of the empirical distributions.
    }
    \label{fig:curvature_pdfs}
\end{figure}

We first study the dependence of ensemble means $\langle (H_{\alpha,\beta})_{ij} \rangle$ ($i,j\in\{1,2\}$) of elements of the dimension-reduced Hessian $H_{\alpha,\beta}$ on the number of samples $S$. According to Eq.~\eqref{eq:expected_hessian}, the diagonal elements of $\mathds{E}[H_{\alpha,\beta}]$ are proportional to the mean curvature of the original high-dimensional space and are thus useful to examine curvature properties of high-dimensional loss functions. Figure~\ref{fig:hessians_random}(a) shows the convergence of the ensemble means $\langle (H_{\alpha,\beta})_{ij} \rangle$ toward the expected values $\mathds{E}[(H_{\alpha,\beta})_{ij}]$ as a function of $S$. Note that $\mathds{E}[(H_{\alpha,\beta})_{ij}]=0$ for all $i,j$. For a few dozen loss projections, the deviations of some of the ensemble means from the corresponding expected values reach values larger than 20. A relatively large number of loss projections $S$ between $10^3$--$10^4$ is required to keep these deviations at values that are smaller than about 2--4. The solid black and red lines in Fig.~\ref{fig:hessians_random}(b), respectively, show the ensemble means $\langle \kappa_\pm^{\alpha,\beta} \rangle$ and
\begin{equation}
\langle \tilde{\kappa}_\pm^{\alpha,\beta} \rangle = \frac{1}{2}\left(\langle A\rangle+\langle C\rangle \pm \sqrt{4 \langle B\rangle ^2+(\langle A\rangle-\langle C\rangle)^2}\right)
\label{eq:kappa_tilde_sample}
\end{equation}
as a function of $S$. Since $\mathds{E}[A]=\mathds{E}[C]={(H_\theta)^i}_i$ and $\mathds{E}[B]=0$ [see Eq.~\eqref{eq:expected_hessian}], we have that $\langle \tilde{\kappa}_\pm^{\alpha,\beta} \rangle=\bar{\kappa}^{\alpha,\beta}$ in the limit $S\rightarrow\infty$. In the current example, the ensemble means $\langle \tilde{\kappa}_\pm^{\alpha,\beta} \rangle$ thus approach $\bar{\kappa}^{\alpha,\beta}=0$ for large numbers of samples $S$, represented by the dashed red lines in Fig.~\ref{fig:hessians_random}(b). The ensemble means $\langle \kappa_{\pm}^{\alpha,\beta}\rangle $ converge towards values of opposite sign, indicating a saddle point.

For a sample size of $S=10^4$, we show the distribution of the principal curvatures $\kappa_\pm^{\alpha,\beta}$ in Fig.~\ref{fig:curvature_pdfs}(a). We observe that the distributions are plausibly Gaussian. We also calculate the probability $\Pr(\kappa_+^{\alpha,\beta}\kappa_-^{\alpha,\beta}>0)$ that the critical point in the lower-dimensional, random projection does not appear as a saddle (\ie, $\kappa_+^{\alpha,\beta}\kappa_-^{\alpha,\beta}>0$). For the example shown in Fig.~\ref{fig:curvature_pdfs}(a), we find that $\Pr(\kappa_+^{\alpha,\beta}\kappa_-^{\alpha,\beta}>0)\approx 0.3$. That is, in about 30\% of the simulated projections, the lower-dimensional loss landscape wrongly indicates that it does not correspond to a saddle.

Our first example, which is based on the loss function \eqref{eq:loss_symmetric}, shows that the principal curvatures in the lower-dimensional representation of $L(\theta)$ may capture the saddle behavior in the original space if one computes ensemble means $\langle \kappa_{\pm}^{\alpha,\beta}\rangle$ in the lower-dimensional space [see Fig.~\ref{fig:hessians_random}(b)]. However, if one first calculates ensemble means of the elements of the dimension-reduced Hessian $H_{\alpha,\beta}$ to infer $\langle \tilde{\kappa}_\pm^{\alpha,\beta} \rangle$, the loss landscape appears to be flat in this example. We thus conclude that different ways of computing ensemble means (either before or after calculating the principal curvatures) may lead to different results with respect to the ``flatness'' of a dimension-reduced loss landscape. 

\subsubsection{Unequal number of curvature directions}
\label{sec:unequal_number_curvature}
In the next example, we will show that random projections cannot identify certain saddle points regardless of the underlying averaging process. We consider the loss function
\begin{equation}
L(\theta)=\frac{1}{2}\theta_{2n+1}\left(\sum_{i=1}^{\tilde{n}}\theta_i^2-\sum_{i=\tilde{n}+1}^{2n}\theta_{i}^2\right)\,,\quad n\in\mathbb{Z}_{+}\,,n<\tilde{n}\leq 2 n\,,
\label{eq:loss_asymmetric}
\end{equation}
where we use the convention $\sum_{i=a}^{b} (\cdot) = 0$ if $a>b$. The Hessian at the critical point $(\theta^*_1,\dots,\theta^*_{2n},\theta^*_{2n+1})=(0,\dots,0,1)$ is 
\begin{equation}
H_\theta=\mathrm{diag}(\underbrace{1,\dots,1}_{\tilde{n}~\mathrm{times}},\underbrace{-1,\dots,-1}_{2n-\tilde{n}~\mathrm{times}},0)\,.
\label{eq:emp_hessian_2}
\end{equation}
As in the previous example, the critical point is again a saddle, but the mean curvature is ${H=2(\tilde{n}-n)/N>0}$. In the following numerical experiments, we set $n=500$ and $\tilde{n}=800$. Figure~\ref{fig:hessians_random}(c) shows that the ensemble means $\langle (H_{\alpha,\beta})_{ij}\rangle$ converge towards the expected value $\mathds{E}[(H_{\alpha,\beta})_{ij}]$ as the number of samples increases. We again observe that a relatively large number of random loss projections $S$ between $10^3$ and $10^4$ is required to keep the deviations of ensemble means from their corresponding expected values small. Because of the dominance of positive principal curvatures $\kappa_i^\theta$ in the original space, the corresponding ensemble means of principal curvatures (\ie, $\langle \kappa_{\pm}^{\alpha,\beta} \rangle$, $\langle \tilde{\kappa}_\pm^{\alpha,\beta}\rangle$) in the lower-dimensional representation approach positive values [see Fig.~\ref{fig:hessians_random}(d)]. The distribution of $\kappa_\pm^{\alpha,\beta}$ indicates that the probability of observing a saddle in the lower-dimensional loss landscape is vanishingly small [see Fig.~\ref{fig:curvature_pdfs}(b)]. In this second example, both ways of computing ensemble means, before and after calculating the lower-dimensional principal curvatures, mistakenly suggest that the saddle in the original space is a minimum in dimension-reduced space.

To summarize, for both loss functions \eqref{eq:loss_symmetric} and \eqref{eq:loss_asymmetric}, the saddle point $\theta^*=(0,\dots,0,1)$ in the original loss function $L(\theta)$ is often misrepresented in lower-dimensional representations $L(\theta+\alpha \eta+\beta\delta)$ if random directions are used. Depending on (i) the employed curvature measure and (ii) the index of the underlying Hessian $H_\theta$ in the original space, the saddle $\theta^*=(0,\dots,0,1)$ may appear erroneously as either a minimum, maximum, or an almost flat region.

If the critical point were a minimum or maximum (\ie, a critical point associated with a positive definite or negative definite Hessian $H_\theta$), it would be correctly represented in the corresponding expected random projection because the sign of its principal curvature $\bar{\kappa}^{\alpha,\beta}$, which is proportional to the sum of all eigenvalues $\kappa_i^\theta$ of the Hessian $H_{\theta}$ in the original loss space, would be equal to the sign of the principal curvatures $\kappa_i^\theta$. However, such points are scarce in high-dimensional loss spaces~\cite{DBLP:conf/nips/DauphinPGCGB14,baldi1989neural}.

Finally, because of the confounding factors associated with the inability of random projections to correctly identify saddle information, it does not appear advisable to use ``flatness'' around a critical point in a lower-dimensional random loss projection as a measure of generalization error~\cite{DBLP:conf/nips/Li0TSG18}.
\subsubsection{Curvature-based Hessian trace estimation}
\label{sec:hessian_trace}
\begin{figure}[t]
    \centering
    \includegraphics[width=.49\textwidth]{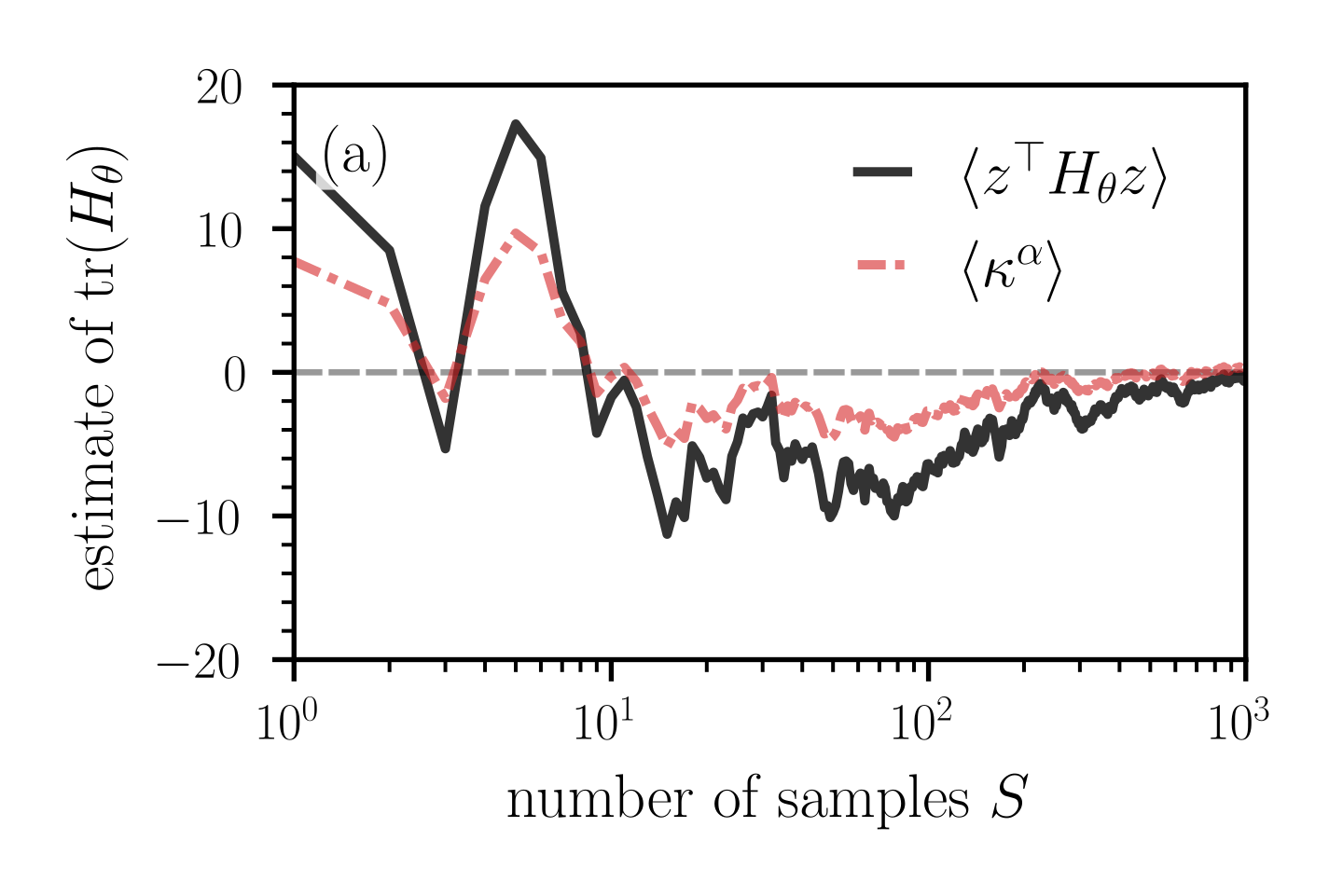}
    \includegraphics[width=.49\textwidth]{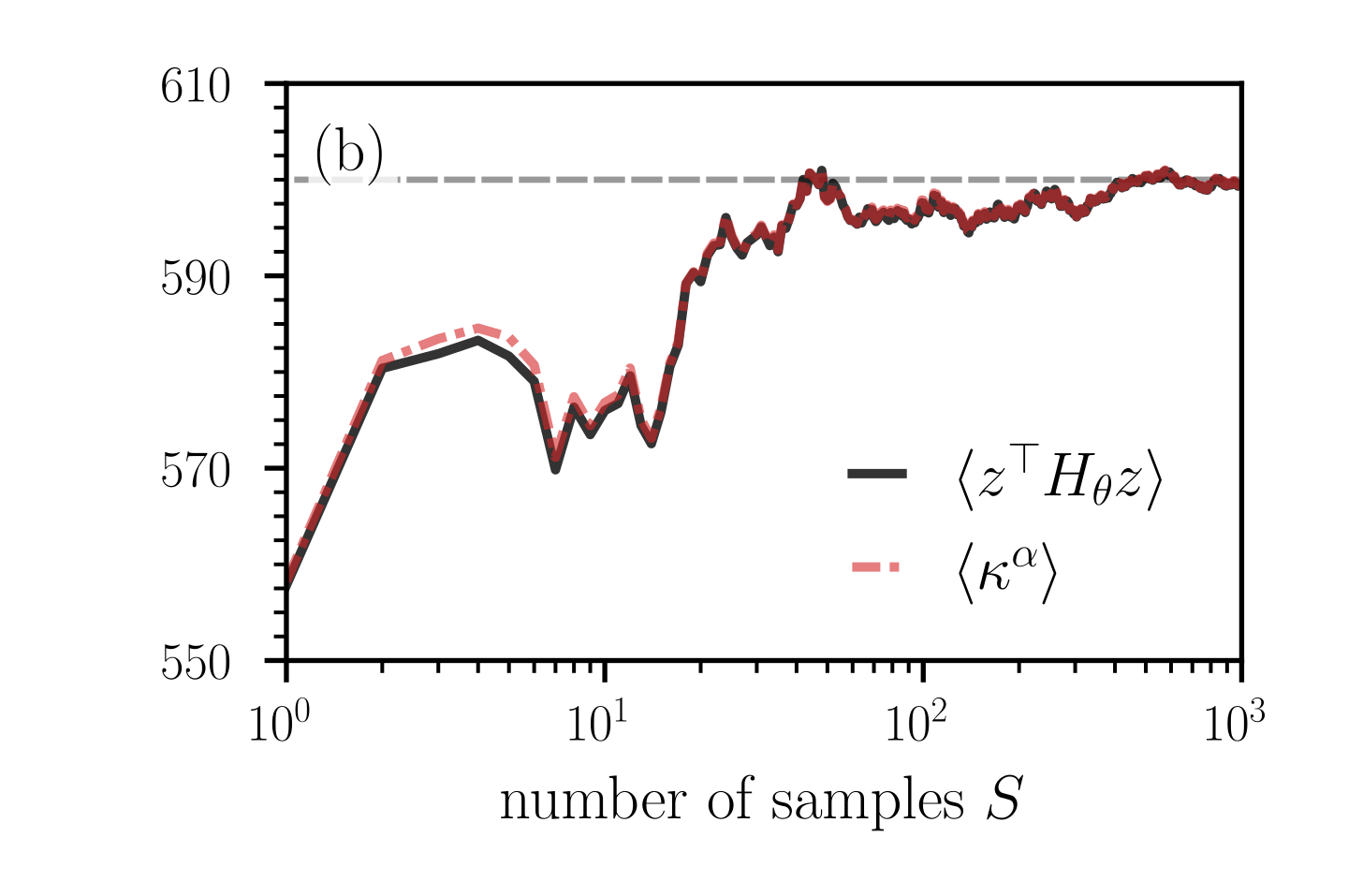}
    \caption{Estimating the trace of the Hessian $H_\theta$. In panels (a) and (b), the loss functions are given by Eqs.~\eqref{eq:loss_symmetric} and \eqref{eq:loss_asymmetric}, respectively. We evaluate the corresponding Hessians \eqref{eq:emp_hessian_1} and \eqref{eq:emp_hessian_2} at the saddle point $\theta^*=(\theta^*_1,\dots,\theta^*_{2n},\theta^*_{2n+1})=(0,\dots,0,1)$. In both loss functions, we set $n=500$ and in loss function~\eqref{eq:loss_asymmetric} we set $\tilde{n}=800$. Solid black and dash-dotted red lines represent Hutchinson [$\langle z^\top H_\theta z \rangle$; see Eq.~\eqref{eq:hutchinson}] and curvature-based ($\langle \kappa^\alpha\rangle$) estimates of $\mathrm{tr}(H_\theta)$, respectively. We compute ensemble means $\langle \cdot \rangle$ as defined in Eq.~\eqref{eq:ensemble_average} for different numbers of random projections $S$. The trace estimates in panels (a) and (b), respectively, converge towards the true trace values $\mathrm{tr}(H_\theta)=0$ and $\mathrm{tr}(H_\theta)=600$ that are indicated by dashed grey lines. In both methods, the same random vectors with elements that are distributed according to a standard normal distribution are used. For the curvature-based estimation of $\mathrm{tr}(H_\theta)$, we perform least-square fits of $L(\theta^*+\alpha\eta)$ over an interval $\alpha\in[-0.05,0.05]$.
    }
    \label{fig:trace_estimation}
\end{figure}
In accordance with Sec.~\ref{sec:hessian_trace_estimates}, we now use the loss functions~\eqref{eq:loss_symmetric} and \eqref{eq:loss_asymmetric} to compare the convergence behavior between the original Hutchinson method~\eqref{eq:hutchinson} and the curvature-based trace estimation~\eqref{eq:curvature_relations}. Instead of two random directions, we use one random Gaussian direction $\eta$ and perform least-squares fits for 50 equidistant values of $\alpha$ in the interval $[-0.05,0.05]$ to extract estimates of $\mathrm{tr}(H_\theta)$ from $L(\theta^*+\alpha\eta)$. We use the same random Gaussian directions in Hutchinson's method.

Figure~\ref{fig:trace_estimation} shows how the Hutchinson and curvature-based trace estimates converge towards the true trace values, 0 for the loss function~\eqref{eq:loss_symmetric} and 600 for the loss function \eqref{eq:loss_asymmetric} with $n=500$ and $\tilde{n}=800$. Given that we use the same random vectors in both methods, their convergence behavior towards the true trace value is similar. 

With the curvature-based method, one can produce Hutchinson-type trace estimates without computing Hessian-vector products. However, it requires the user to specify an appropriate interval for $\alpha$ so that the mean curvature can be properly estimated in a quadratic-approximation regime. It also requires a sufficiently large number of points in this interval. Therefore, it may be less accurate than the original Hutchinson method.
\subsection{Hessian directions}
\label{sec:hessian_directions}
\begin{figure}
    \centering
    \includegraphics[width=.75\textwidth]{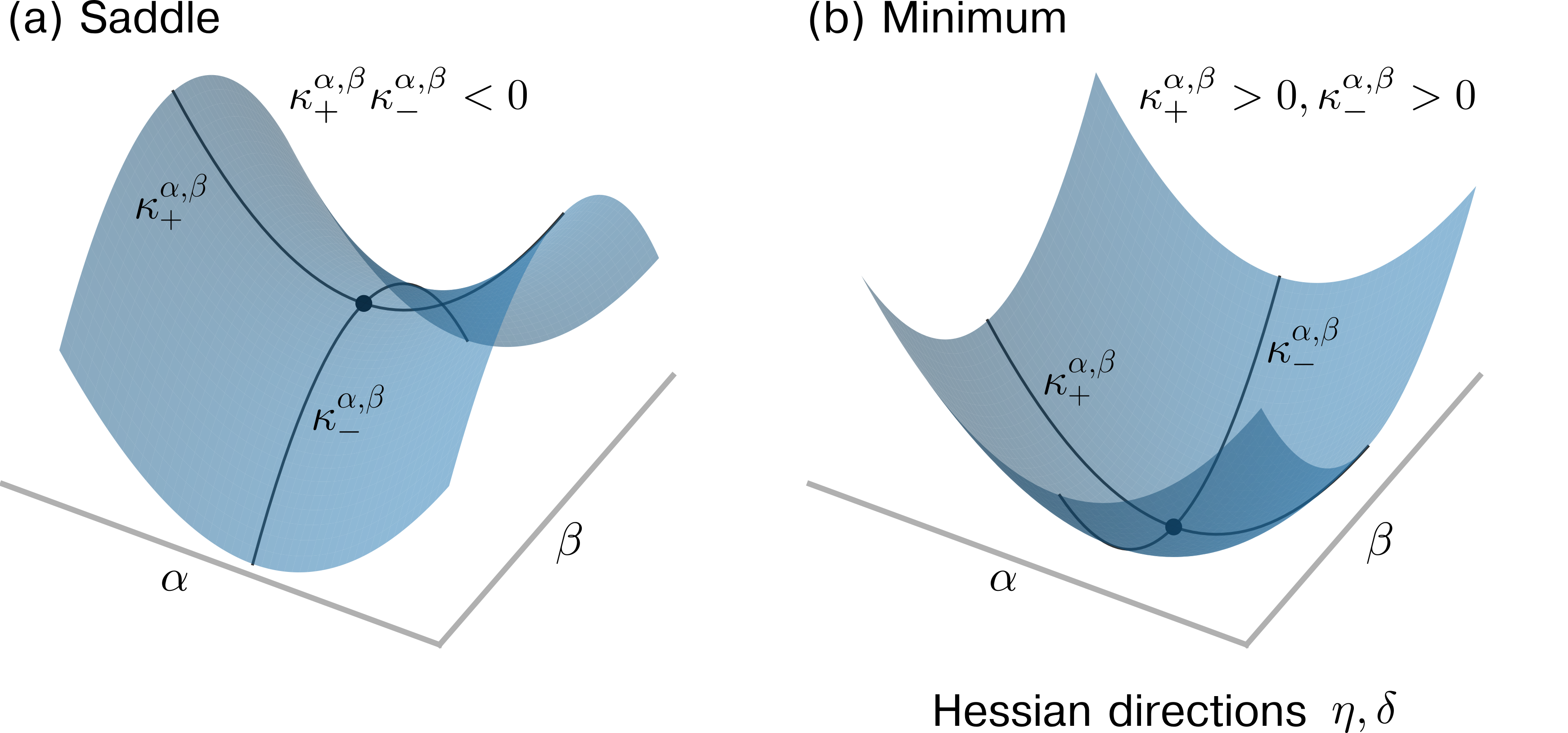}\\[15pt]
    \includegraphics[width=.75\textwidth]{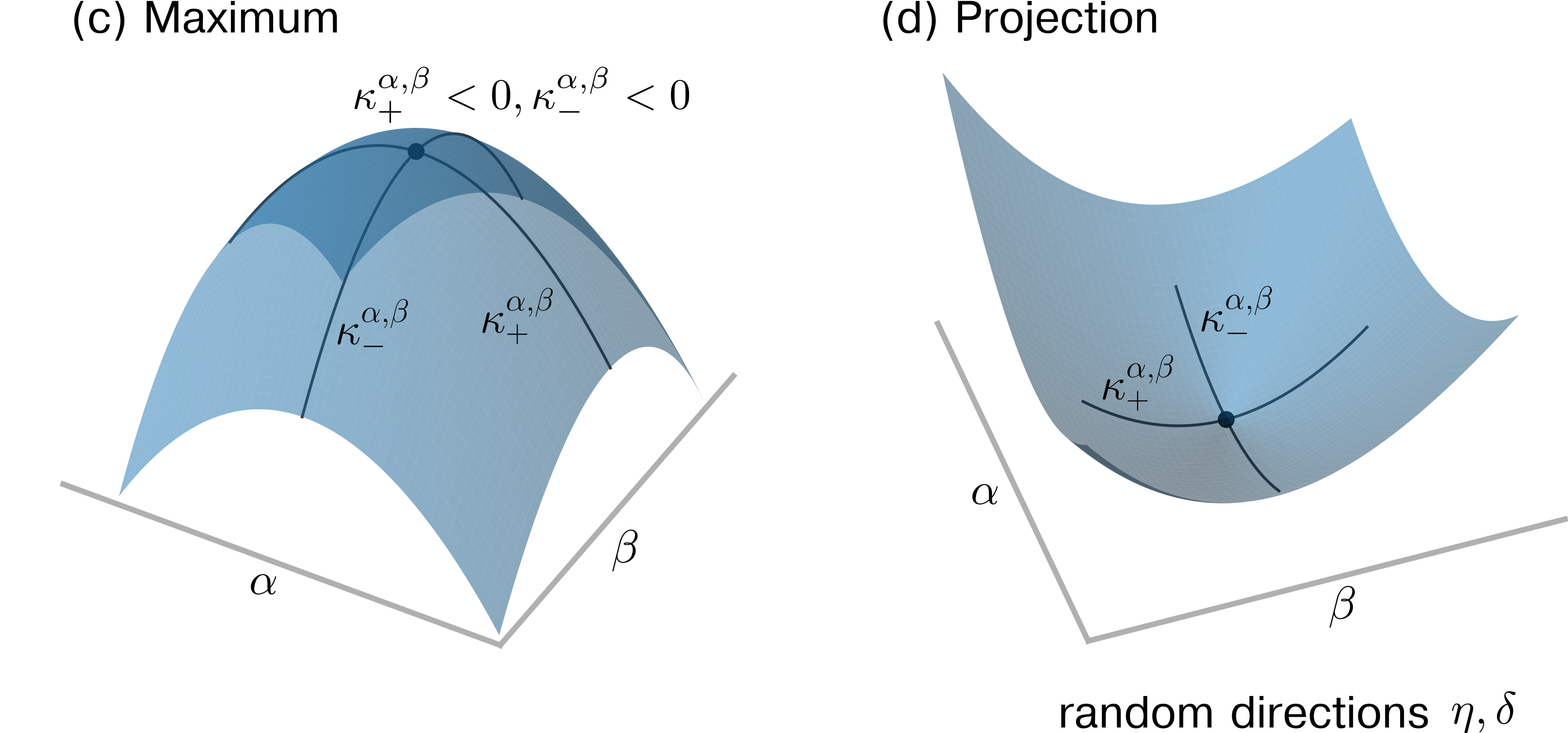}
    \caption{Dimensionality-reduced loss $L(\theta^*+\alpha \eta+\beta\delta)$ of Eq.~\eqref{eq:loss_asymmetric} with $n=900,\tilde{n}=1000$ for different directions $\eta,\delta$. (a--c) The directions $\eta,\delta$ correspond to eigenvectors of the Hessian $H_\theta$ of Eq.~\eqref{eq:loss_asymmetric}. If the eigenvalues associated with $\eta,\delta$ have different signs, the corresponding loss landscape is a saddle as depicted in panel (a). If the eigenvalues associated with $\eta,\delta$ have the same sign, the corresponding loss landscape is either a minimum (both signs are positive) as shown in panel (b) or a maximum (both signs are negative) as shown in panel (c). Because there is an excess of $\tilde{n}-n=100$ positive eigenvalues in $H_\theta$, a projection onto a dimension-reduced space that is spanned by the random directions $\eta,\delta$ is often associated with an apparently convex loss landscape. An example of such an apparent minimum is shown in panel (d).
    }
    \label{fig:loss_projection}
\end{figure}
Given the described shortcomings of random projections incorrectly identifying saddles of high-dimensional loss functions, we suggest to use Hessian directions (\ie, the eigenbasis of $H_\theta$) as directions $\eta,\delta$ in $L(\theta^*+\alpha \eta + \beta \delta)$.

For Eq.~\eqref{eq:loss_asymmetric} with $n=900,\tilde{n}=1000$, we show projections along different Hessian directions in Fig.~\ref{fig:loss_projection}(a--c). We observe that different Hessian directions indicate different types of critical points in dimension-reduced space. If the eigenvalues associated with the Hessian directions $\eta,\delta$ have different signs, the corresponding lower-dimensional loss landscape is a saddle [see Fig.~\ref{fig:loss_projection}(a)]. If both eigenvalues have the same sign, the loss landscape is either a minimum [see Fig.~\ref{fig:loss_projection}(b): both signs are positive] or it is a maximum [see Fig.~\ref{fig:loss_projection}(c): both signs are negative]. If one uses a random projection instead, the resulting lower-dimensional loss landscape often appears to be a minimum in this example [see Fig.~\ref{fig:loss_projection}(d)]. To quantify the proportion of random projections that correctly identify the saddle with $n=900,\tilde{n}=1000$, we generated 10,000 realizations of $\kappa_{\pm}^{\alpha,\beta}$ [see Eq.~\eqref{eq:kapp_alpha_beta}]. We find that the signs of $\kappa_{\pm}^{\alpha,\beta}$ were different in only about 0.5\% of all simulated realizations. That is, in this example the principal curvatures $\kappa_{\pm}^{\alpha,\beta}$ indicate a saddle in only about 0.5\% of the studied projections. 

In accordance with related works that use Hessian information~\cite{yao2020pyhessian,liao2021hessian}, our proposal uses {\em Hessian-vector products} (HVPs). We first compute the largest-magnitude eigenvalue and then use an annihilation method~\cite{burden2015numerical} to compute the second largest-magnitude eigenvalue of opposite sign (see Algorithm~\ref{alg:hessian_directions}). The corresponding eigenvectors are the dominant Hessian directions. Other deflation techniques can be used to find additional Hessian directions.

We employ HVPs to compute Hessian directions without an explicit representation of $H_\theta$ using the identity
\begin{equation}
\nabla_\theta [(\nabla_\theta L)^\top v]=(\nabla_\theta \nabla_\theta L) v +(\nabla_\theta L)^\top \nabla_\theta v= H_\theta v\,.
\end{equation}
In the first step, the gradient $(\nabla_\theta L)^\top$ is computed using reverse-mode autodifferentiation (AD) to then compute the scalar product $[(\nabla_\theta L)^\top v]$. In the second step, we again apply reverse-mode AD to the computational graph associated with the scalar product $[(\nabla_\theta L)^\top v]$. Because the vector $v$ does not depend on $\theta$ (\ie, $\nabla_\theta v=0$), the result is $H_\theta v$. One may also use forward-mode AD in the second step to provide a more memory efficient implementation.
\begin{algorithm}[tb]
   \caption{Compute dominant Hessian directions}
   \label{alg:hessian_directions}
\begin{algorithmic}[1]
\STATE{$L_1=\mathrm{LinearOperator((N,N),~matvec=hvp)}$}\\
{\footnotesize \em initialize linear operator for HVP calculation}
\STATE{eigval1, eigvec1 = solve\_lm\_evp($L_1$)}\\
{\footnotesize \em compute largest-magnitude eigenvalue and corresponding eigenvector associated with operator $L_1$}
\STATE{shifted\_hvp(vec) = hvp(vec) - eigval1*vec}\\
{\footnotesize \em define shifted HVP}
\STATE{$L_2=\mathrm{LinearOperator((N,N),~matvec=shifted\_hvp)}$}\\
{\footnotesize \em initialize linear operator for shifted HVP calculation}
\STATE{eigval2, eigvec2 = solve\_lm\_evp($L_2$)}\\
{\footnotesize \em compute largest-magnitude eigenvalue and corresponding eigenvector associated with operator $L_2$}
\STATE{eigval2 += eigval1}
\IF{eigval1 $>=$ 0}
\STATE{maxeigval, maxeigvec = eigval1, eigvec1}
\STATE{mineigval, mineigvec = eigval2, eigvec2}
\ELSE
\STATE{maxeigval, maxeigvec = eigval2, eigvec2}
\STATE{mineigval, mineigvec = eigval1, eigvec1}
\ENDIF
\STATE {\bfseries return:} {maxeigval, maxeigvec, mineigval, mineigvec}
\end{algorithmic}
\end{algorithm}
\subsubsection{Image classification}
\begin{figure}
    \centering
    \includegraphics[width=.95\textwidth]{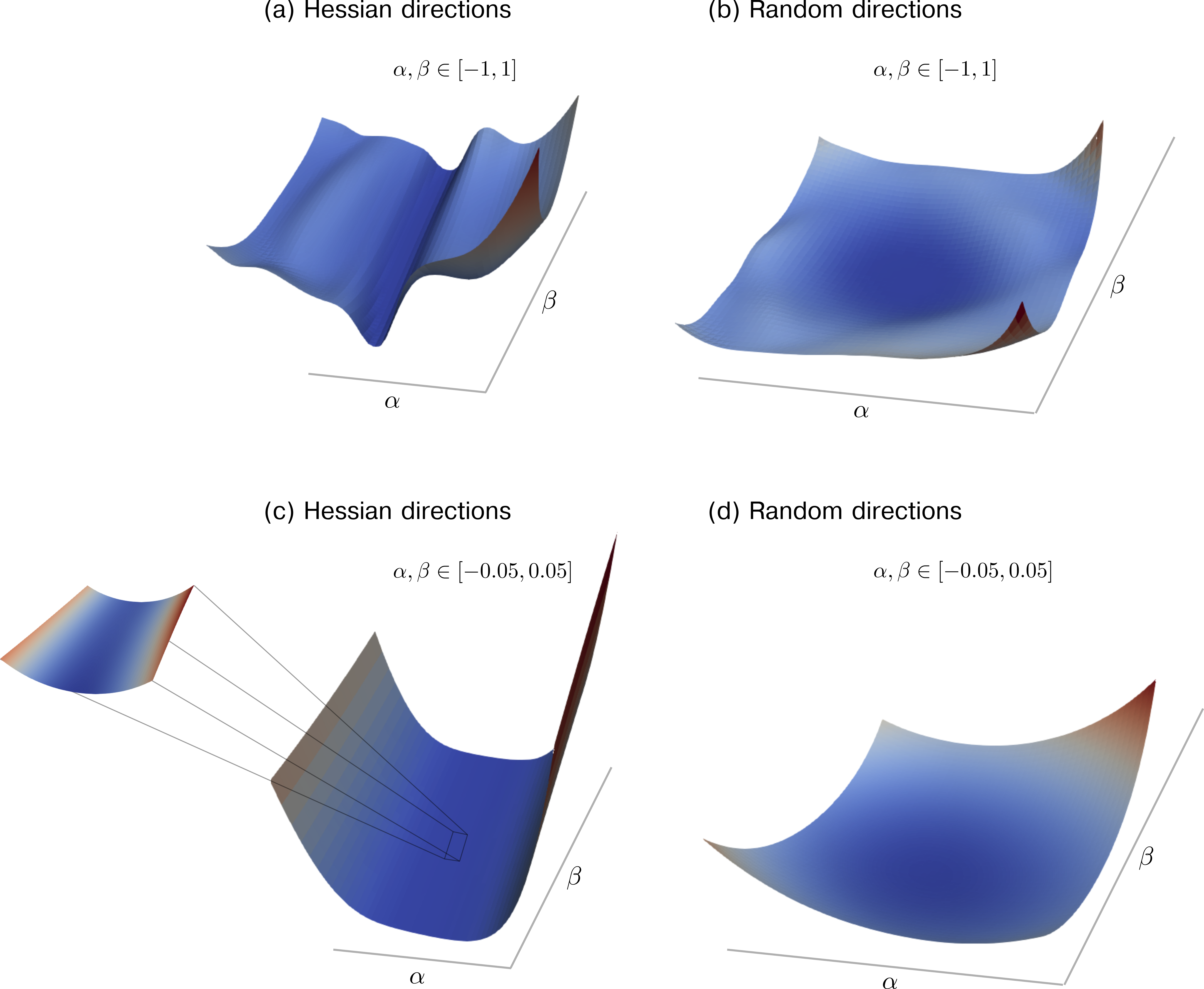}
    \caption{Loss landscape projections for ResNet-56. (a,c) The projection directions $\eta,\delta$ are given by the eigenvectors associated with the largest and smallest eigenvalues of the Hessian $H_\theta$, respectively. The zoomed inset in panel (c) shows the loss landscape for $(\alpha,\beta)\in[-0.01,0.005]\times[-0.05,0.05]$. We observe a decreasing loss along the negative $\beta$-axis. (b,d) The projection directions $\eta,\delta$ are given by random vectors. The domains of $(\alpha,\beta)$ in panels (a,b) and (c,d) are $[-1,1]\times[-1,1]$ and $[-0.05,0.05]\times[-0.05,0.05]$, respectively. All shown cross-entropy loss landscapes are based on evaluating the CIFAR-10 training dataset that consists of 50,000 images.}
    \label{fig:resnet_56}
\end{figure}
As an illustration of a loss landscape of an ANN employed in image classification, we focus on the ResNet-56 architecture that has been trained as detailed in Ref.~\refcite{DBLP:conf/nips/Li0TSG18} on {CIFAR-10} using stochastic gradient descent (SGD) with Nesterov momentum. The number of parameters of this ANN is 855,770. The training and test losses at the local optimum found by SGD are ${9.20\times 10^{-4}}$ and 0.29, respectively; the corresponding accuracies are 100.00 and 93.66, respectively. 

In accordance with Ref.~\refcite{DBLP:conf/nips/Li0TSG18}, we apply filter normalization to random directions. This and related normalization methods are often employed when generating random projections. One reason is that simply adding random vectors to parameters of a neural network loss function (or parameters of other functions) does not consider the range of parameters associated with different elements of that function. As a result, random perturbations may be too small or large to properly resolve the influence of certain parameters on a given function.

When calculating Hessian directions, we are directly taking into account the parameterization of the underlying functions that we want to visualize. Therefore, there is no need for an additional rescaling of different parts of the perturbation vector. Still, reparameterizations of a neural network can result in changes of curvature properties (see, \eg, Theorem 4 in Ref.~\refcite{DBLP:conf/icml/DinhPBB17}).

Figure~\ref{fig:resnet_56} shows the two-dimensional projections of the loss function (cross entropy loss) around a local critical point. The smallest and largest eigenvalues are $-16.4$ and $5007.9$, respectively. This means that the found critical point is a saddle with a maximum negative curvature that is more than two orders of magnitude smaller than the maximum positive curvature at that point. The saddle point is clearly visible in Fig.~\ref{fig:resnet_56}(a,c). We observe in the zoomed inset in Fig.~\ref{fig:resnet_56}(c) that the loss decreases along the negative $\beta$-axis.

If random directions are used, the corresponding projections indicate that the optimizer converged to a local minimum and not to a saddle point [see Fig.~\ref{fig:resnet_56}(b,d)]. Overall, the ResNet-56 visualizations that we show in Fig.~\ref{fig:resnet_56} exhibit structural similarities to those that we generated using the simple loss model \eqref{eq:loss_asymmetric} [see Fig.~\ref{fig:loss_projection}(a,d)].
\subsubsection{Function approximation}
\begin{figure}
    \centering
    \includegraphics[width=\textwidth]{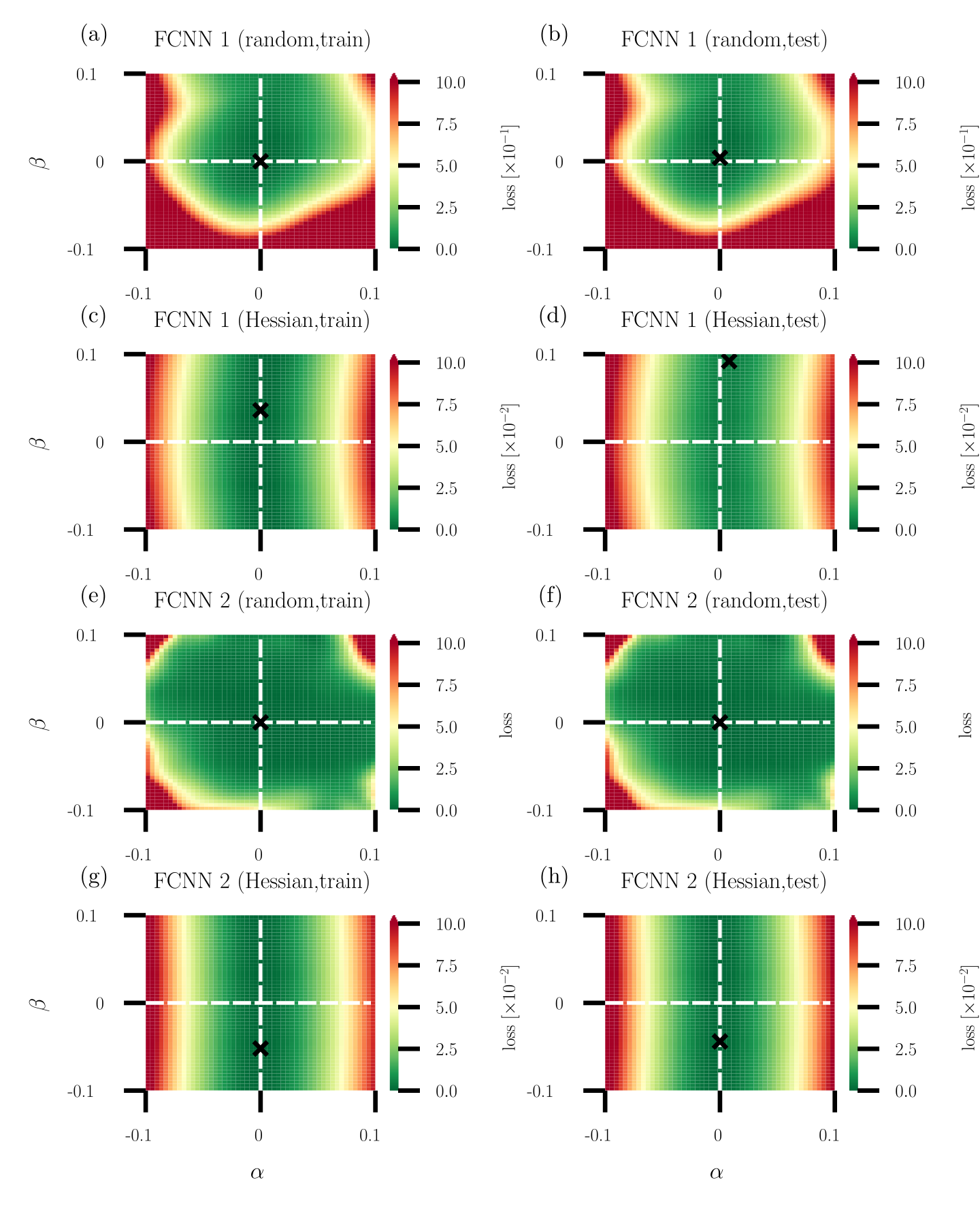}
    \caption{Heatmaps of the mean squared error (MSE) loss along random and dominant Hessian directions for a function-approximation task. (a--d) Loss heatmaps for a fully connected neural network (FCNN) with 2 layers and 100 ReLU activations per layer [(a,c): training data, (b,d): test data]. (e--h) Loss heatmaps for an FCNN with 10 layers and 100 ReLU activations per layer [(e,g): training data, (f,h): test data]. Random directions are used in panels (a,b,e,f) while Hessian directions are used in panels (c,d,g,h). Green and red regions indicate small and large mean squared error (MSE) loss values, respectively. Black crosses indicate the positions of loss minima in the shown domain. Both neural networks, FCNN 1 and FCNN 2, are trained to approximate the smooth one-dimensional function \eqref{eq:func_approx}.}
    \label{fig:train_test_function_approximation}
\end{figure}
As another example, we compare loss function visualizations that are based on random and Hessian directions in a function-approximation task. In accordance with Ref.~\refcite{adcock2021gap}, we consider the smooth one-dimensional function
\begin{equation}
    f(x)=\log(\sin(10x)+2)+\sin(x)\,,
\label{eq:func_approx}
\end{equation}
where $x\in [-1,1)$. To approximate $f(x)$, we use two fully connected neural networks (FCNNs) with 2 and 10 layers, respectively. Each layer has 100 ReLU activations. The numbers of parameters of the 2 and 10 layer architectures are 20,501 and 101,301, respectively. The training data is based on 50 points that are sampled uniformly at random from the interval $[-1,1)$. We train both ANNs using a mean squared error (MSE) loss function and SGD with a learning rate of 0.1. The 2 and 10 layer architectures are respectively trained for 100,000 and 50,000 epochs to reach loss values of less than $10^{-4}$. The best model was saved and evaluated by calculating the MSE loss for 1,000 points that were sampled uniformly at random from the interval $[-1,1)$.\footnote{We present an animation illustrating the evolution of random and Hessian loss projections during training at \url{https://vimeo.com/787174225}.}

Figure~\ref{fig:train_test_function_approximation} shows heatmaps of the training and test loss landscapes of both ANNs along random and dominant Hessian directions. Black crosses in Fig.~\ref{fig:train_test_function_approximation} indicate loss minima. In line with the previous example, which focused on an image classification ANN, we observe for both FCNNs that random projections are associated with loss values that increase along both directions $\delta,\eta$ and for both training and test data [see Fig.~\ref{fig:train_test_function_approximation}(a,b,e,f)]. For these projections, we find that the loss minima are very close to or at the origin of the loss space. The situation is different in the loss projections that are based on Hessian directions. Figure~\ref{fig:train_test_function_approximation}(c) shows that the loss minimum for the 2-layer FCNN is not located at the origin but at $(\alpha,\beta)\approx(0,0.04)$. 

We find that the value of the training loss at that point is more than 9\% smaller than the smallest training loss found in a random projection plot. The corresponding test loss is about 1\% smaller than the test loss associated with the smallest training loss in the random projection plot. For the 10-layer FCNN, the smallest training loss in the Hessian direction projection is more than 26\% smaller than the smallest training loss in the randomly projected loss landscape [see Fig.~\ref{fig:train_test_function_approximation}(e,g)]. The corresponding test loss in the Hessian direction plot is about 6\% smaller than the corresponding test loss minimum in the random direction plot [see Fig.~\ref{fig:train_test_function_approximation}(f,h)]. Notice that both the training and test loss minima in the random direction heatmaps in Fig.~\ref{fig:train_test_function_approximation}(e,f) are located at the origin while they are located at $(\alpha,\beta)\approx(0,-0.05)$ in the Hessian direction heatmaps in Fig.~\ref{fig:train_test_function_approximation}(g,h).

These results show that it is possible to improve the training loss values along the dominant negative curvature direction. Smaller loss values may also be achievable with further gradient-based training iterations, so one has to consider tradeoffs between gradient
and curvature-based optimization methods.
\section{Conclusions}
\label{sec:conclusions}
Over several decades, research at the intersection of statistical mechanics, neuroscience, and computer science has significantly enhanced our understanding of information processing in both living systems and machines. With the increasing computing power, numerous learning algorithms conceived in the latter half of the 20th century have found widespread application in tasks such as image classification, natural language processing, and anomaly detection. 

The Ising model, one of the most thoroughly studied models in statistical mechanics, shares close connections with ANNs such as Hopfield networks and Boltzmann machines. In the first part of this chapter, we described these connections and provided an illustrative example of a Boltzmann machine that learned to generate new Ising configurations based on a set of training samples.

Despite decades of active research in artificial intelligence and machine learning, providing mechanistic insights into the representational capacity and learning dynamics of modern ANNs remains challenging due to their high dimensionality and complex, non-linear structure. Visualizing the loss landscapes associated with ANNs poses a particular challenge, as one must appropriately reduce dimensionality to visualize landscapes in two or three dimensions. To put it in the words of Ker-Chau Li, ``There are too many directions to project a high-dimensional data set and unguided plotting can be time-consuming and fruitless''. One approach is to employ random projections. However, this method often inaccurately depicts saddle points in high-dimensional loss landscapes as apparent minima in low-dimensional random projections.

In the second part of this chapter, we discussed various curvature measures that can be employed to characterize critical points in high-dimensional loss landscapes. Additionally, we outlined how Hessian directions (\ie, the eigenvectors of the Hessian at a critical point) can inform a more structured approach to projecting high-dimensional functions onto lower-dimensional representations.

While much of contemporary research in machine learning is driven by practical applications, incorporating theoretical contributions rooted in concepts from statistical mechanics and related disciplines holds promise for continuing to guide the development of learning algorithms.
\section*{Acknowledgements}
LB acknowledges financial support from hessian.AI and the ARO through grant W911NF-23-1-0129.
\bibliographystyle{ws-rv-van}
\bibliography{ws-rv-sample,refs}


\end{document}